# Remotely programming the weights of a spintronic neural network by a radiofrequency broadcast signal

M. Menshawy[1], D. Sanz-Hernández[1], L. Mazza[2], V. Puliafito[2], G. Finocchio[3], A. Jenkins[4], R. Ferreira[4], L. Benetti[4], J. Grollier[1*], and F.A. Mizrahi[1*]

1- Laboratoire Albert Fert, CNRS, Thales, Université Paris-Saclay, Palaiseau, France
2- Department of Electrical and Information Engineering, Politecnico di Bari, 70126, Bari, Italy
3- Department of Mathematical and Computer Sciences, Physical Sciences and Earth Sciences, University of Messina, Messina, Italy
4- International Iberian Nanotechnology Laboratory, Braga, Portugal

* Corresponding authors: julie.grollier@cnrs-thales.fr, frank.mizrahi@cnrs-thales.fr

**Selectively programming a large number of non-volatile synaptic weights without compromising scalability is a key challenge for in-memory computing. Here, we experimentally demonstrate remote programming of synaptic weights in series-connected chains of 11 vortex-based magnetic tunnel junctions using broadcast radiofrequency signals applied through a shared strip line. Because each junction is engineered with a distinct resonance frequency, programming relies on frequency-selective reversal of the vortex-core polarity and therefore does not require individual access lines or selector devices. Reconfiguring the binary states of these chains strongly reshapes the continuous spectral transfer functions that determine the weighted sums performed on frequency-multiplexed RF inputs. This allows compact networks built from such chains to process high-dimensional RF-encoded inputs. We experimentally perform handwritten-digit classification and drone RF-signature identification using a two-chain network comprising only 22 synapses in total. After remote reconfiguration for each task, the spintronic network respectively reaches 93.9 ± 2.3 % and 99.5 ± 1 % test accuracy, on par with the software baseline. Broadcast RF programming thus provides a compact and scalable route to rapidly reconfigurable spintronic neuromorphic hardware.**

In-memory computing constitutes a promising route toward more energy-efficient artificial-intelligence hardware as it reduces data transfer between memory and processing units and allows physical devices to directly perform vector-matrix operations[1–4]. A central challenge, however, is to program the individual non-volatile memory components storing the weights in these dense networks without sacrificing compactness, energy efficiency, or scalability.

Existing approaches face important trade-offs. Passive crossbar arrays require complex biasing schemes across rows and columns, which increase circuit overhead and programming energy[5,6]. As illustrated in Figure 1a, adding a transistor-based selector restores individual control but increases the cell area by factors up to two orders of magnitude in current demonstrations, while also adding local access routing, and complicating dense 3D integration[7–10,5]. In addition, programming and inference typically still rely on distinct peripheral circuits for write driving, sensing and computation. More generally, architectures that rely on individual synapse access preserve programmability, but their wiring overhead makes scaling to dense large networks difficult. This motivates approaches in which many channels can instead be routed through shared physical interconnects.

Among neuromorphic hardware platforms, architectures that exploit frequency multiplexing are especially attractive because many input channels can be routed and processed on shared physical interconnects. This opportunity has been explored for inference in photonic systems based on wavelength-selective microring resonators, microring weight banks and phase-change photonic memories[4,11–16], as well as in spintronic systems including radio-frequency (RF) networks based on magnetic tunnel junctions[17–23]. Yet in all these platforms, frequency multiplexing applies solely to network inputs for inference, while programming relies on individual electrical access to each synapse, limiting scalability. In particular, demonstrations with spintronic RF networks[17–19,21,22,24]. relied on either fixed weights or volatile programming.

Here we introduce a different architecture, illustrated in Fig. 1b, in which chains of magnetic tunnel junctions connected in series are remotely addressed through a single shared capacitively coupled RF strip line for synaptic programming, state readout and RF input delivery. Unlike individually addressed arrays, this scheme does not require per-synapse selector devices nor dedicated access lines. It exploits the frequency selectivity of magnetic tunnel junctions and the microwave-induced reversal of their magnetic state, here their vortex-core polarity. Because each junction is designed with a distinct resonance frequency, a broadcast RF signal can selectively switch or read any synapse within the chain.

Since all 11 junctions in the experimentally realized chain can be independently programmed through this single shared access, the chain can be placed in any of its 2048 binary configurations. Traversing this configuration space strongly reshapes the spectral transfer function of the chain, which we exploit for RF signal classification. During inference, RF inputs are frequency multiplexed, so that the input dimensionality is not constrained by the number of magnetic tunnel junctions in the chain. Using two such chains, we physically implement a 22-synapse network that can be remotely reconfigured between handwritten-digit classification and drone RF-signature identification. After reconfiguration for each task, the experimental network respectively reaches 93.9 ± 2.3 % and 99.5 ± 1 % test accuracy, showing that the same hardware can switch between distinct functions while maintaining task-specific performance on par with the software baseline (Methods). These results identify broadcast RF programming as a scalable route toward compact and rapidly reconfigurable neuromorphic hardware based on spintronic devices.

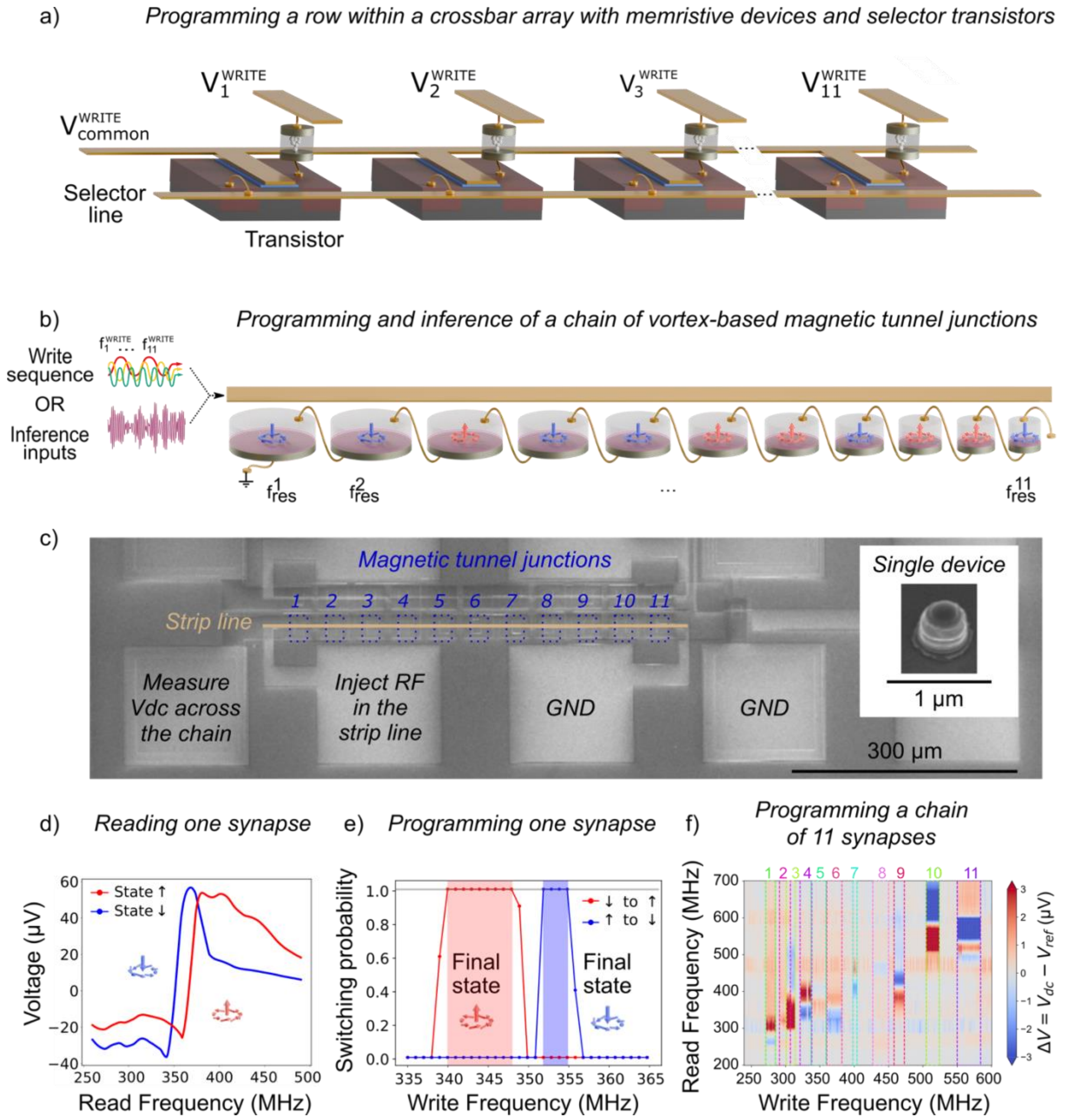

*__Figure 1.__ (a) Schematic of a row of memory devices with selector transistors and access lines within a crossbar array. (b) Schematic of a chain of spintronic synapses programmed by a broadcast RF signal. Both write and inference signals are frequency-multiplexed through a single access to a strip line capacitively coupled to the synapses. (c) Scanning electron microscopy image of the chain of 11 magnetic tunnel junctions with its strip line and contact pads. Insert: single device pillar during fabrication. (d) Reading the synaptic state of a single MTJ: dc voltage $V_{dc}$ across an MTJ versus RF input frequency, at a power of -8 dBm and under a +10 mT perpendicular magnetic field. (e) Programming the synapse: switching probability versus RF input frequency, at a power of 0 dBm under a +10 mT perpendicular magnetic field. The red and blue curves respectively represent switching from down to up and up to down. The red and blue shaded zones respectively represent a final state that is up or down. Probabilities are evaluated over 10 independent trials. (f) Response of the chain $\Delta V$ (voltage difference compared to the reference fully-down state) versus programming frequency and read frequency.*

*The experiment is performed under a -20 mT perpendicular magnetic field. The color scale indicates the voltage difference $\Delta V$ between the chain response after the write attempt and that of the reference state. The colourmap is capped at ± 3 $\mu V$ for readability. The respective switching zones for each of the 11 junctions are highlighted through dotted lines of different colors.*

**RF readout and programming of frequency-selective synapses**

We implement each synapse with a magnetic tunnel junction whose free layer hosts a magnetic vortex ground state (Figure 1b and Methods). In this configuration, the magnetization curls in the device plane around a small (about 20 nm wide) out-of-plane vortex core. The free layer has been optimized for efficient RF dynamics of the vortex core[25]. The synaptic weight is encoded in the polarity of this core, which points either up or down and thus defines a binary state.

Both readout and programming rely on radio-frequency excitation delivered by a metallic strip line placed above the junction and capacitively coupled to it (Figure 1b-c). An RF signal sent through this strip line remotely excites the gyrotropic motion of the vortex core, in which the core orbits in the device plane. Depending on the RF power, this excitation can be used either to read the stored state non-destructively or to switch it.

For readout, we inject a low-power RF signal (-8 dBm) at a frequency close to the gyrotropic resonance. This excitation also induces an RF current through the device, because the strip line is capacitively coupled to the junction. The resulting resistance oscillations associated with the vortex motion mix with this current and generate a dc voltage through the spin-diode effect[26,27]. Under a 10 mT perpendicular magnetic field, the two vortex-core polarities have different resonance frequencies[28,29]. In consequence the measured spin-diode spectrum provides a direct readout of the stored synaptic state, as illustrated by the blue and red curves in Figure 1d for a single MTJ.

For programming, we increase the RF power (between 0 and 13 dBm depending on the frequency, see Methods) and excite the junction close to the gyrotropic resonance. At this larger oscillation amplitude, the vortex core reaches its critical velocity and reverses polarity[30]. This reversal occurs only within a well-defined resonant frequency window. In consequence, the programming process is frequency selective. Figure 1e shows the experimentally measured probability of switching from the up to the down state (blue) and from the down to the up state (red) as a function of the write frequency. Away from resonance, the RF signal leaves the synaptic state unchanged. By contrast, when the write frequency falls within the switching window of one polarity, the junction is programmed reproducibly into either the up state (red region) or the down state (blue region).

We then exploit this frequency selectivity to program chains of 11 serially connected MTJs using a single strip line per chain to read and write the devices. Within each chain, the junction diameters are chosen so as to span distinct gyrotropic resonance frequencies (Methods), allowing the same broadcast RF signal to address different synapses depending on its frequency. Figure 1f maps how the state of a chain changes with programming frequency, starting from the fully down reference state. For each column, we attempt programming at a different frequency as follows. First, we first initialize all 11 MTJs in the down state under a perpendicular field of -300 mT. Then, under a -20 mT field, we apply a high-power RF pulse to the strip-line at the selected write frequency. The write power is adjusted to the selected frequency, as detailed in Methods. Finally, we read the response by applying a low power (-8 dBm) RF signal to the strip-line and measuring the dc voltage across the chain as a function of read frequency. The color scale indicates the voltage difference ΔV between the chain response after the write attempt and that of the reference state. The dashed rectangles highlight the frequency regions programming each MTJ. Using these measurements, we determine the powers and frequencies to program each MTJ. We confirm the reliability of the scheme with 10 independent trials where we reset then program the device (Methods).

**Spectral reconfiguration of synaptic chains**

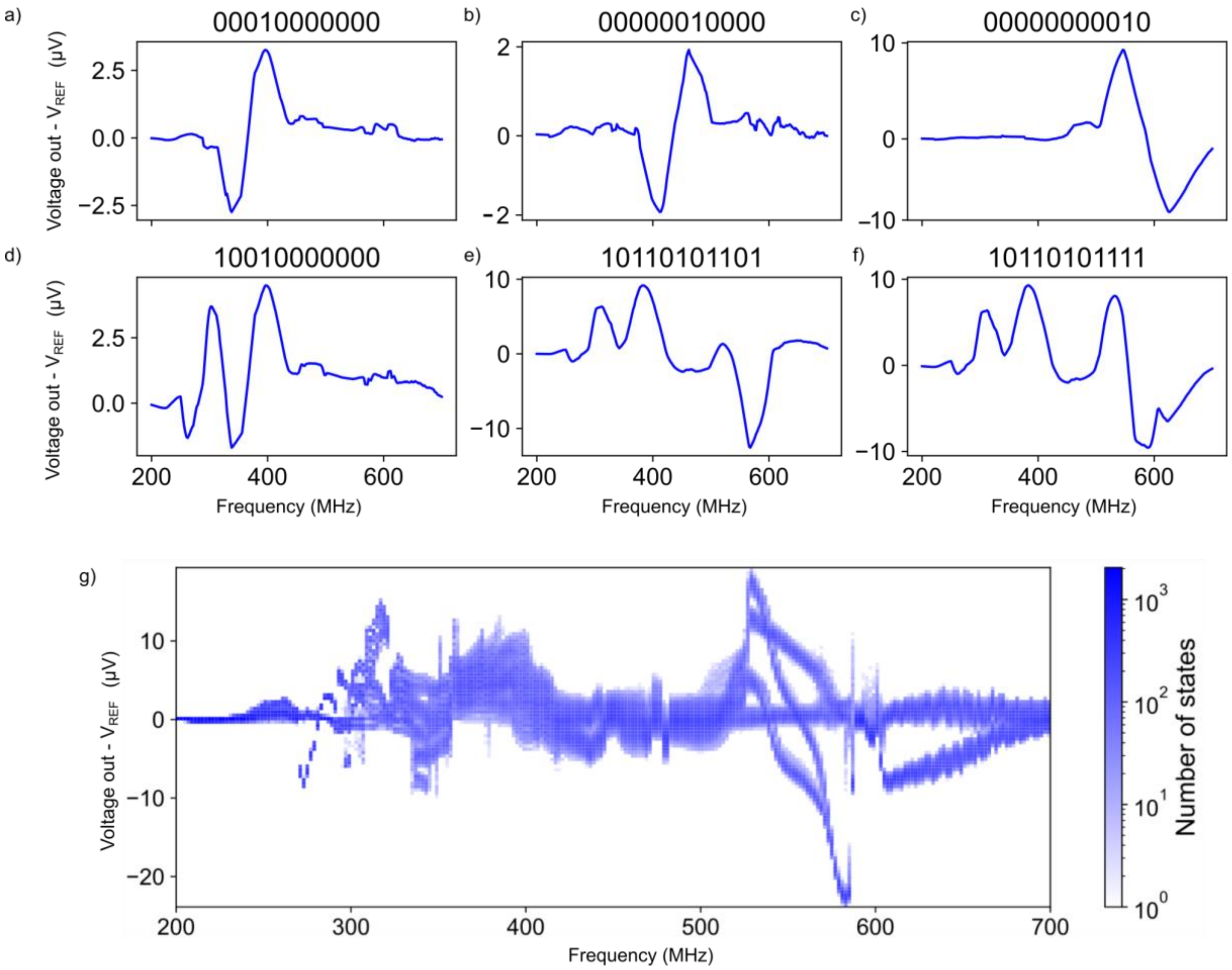

***Figure 2.*** *(a-f) Response of the chain versus input frequency. The response is the voltage difference between the measured configuration and the reference (fully down) state. Each panel corresponds to a different configuration. "1" and "0" respectively refer to up and down states. (g) Density map of the chain response over all 2048 configurations, using 100 voltage bins. For each input frequency, the color scale indicates how many configurations produce a given output voltage.*

By applying sequences of RF write pulses selected from the programming bands identified in Figure 1f, we can set each 11-junction chain into any of its $2^{11}$ = 2048 binary configurations (Methods). Starting from a fully down or fully up reference state, the pulses are applied in a prescribed frequency order to avoid unintended switching events between neighboring write bands. This enables deterministic programming of arbitrary chain states using only a single broadcast RF line per chain.

In a series-connected chain, the individual spin-diode responses of the MTJs add up into a single dc output, so that the chain realizes a weighted sum on frequency-multiplexed RF inputs [17,19,20]. Changing the binary configuration of the chain therefore reshapes its spectral transfer function. The individual MTJ responses are wide enough in frequency to overlap with each other. In consequence, the chain response at each frequency depends on all MTJs, providing a reconfigurable continuous response rather than a fixed set of discrete coefficients.

Figures 2a–f show six representative programmed configurations of the chain. In each panel, the plotted voltage is the difference between the response of the programmed configuration and that of the fully down reference state. Figures 2a–c correspond to configurations in which a single junction is switched to the up state, denoted as "1". In these cases, the chain response is dominated by a single spin-diode feature whose resonance frequency depends on which junction has been reversed. The higher the index of the junction, the higher the frequency of the spin-diode feature. This is due to the fact that the junctions are arranged in decreasing diameter, and thus increasing frequency, order. Figure 2d shows a configuration in which one additional junction is switched to the up state compared with Fig. 2a, resulting in an additional spectral contribution to the overall response. Figures 2e and 2f show configurations with several MTJs in the up state, for which the chain response becomes markedly more structured. Together, these examples show that changing the binary magnetic configuration of the chain reshapes its spectral transfer function in a controlled manner, from simple single-resonance responses to complex multi-feature spectra.

We then evaluate the full functional space accessible to a chain by measuring its response for all 2048 binary configurations. Figure 2g shows the resulting density map: for each input frequency, the blue color scale indicates how many configurations produce a given output voltage. This representation reveals not only the number of accessible responses, but also their span and distribution. At some frequencies, such as around 550 MHz, the output is concentrated onto a few discrete voltage levels, each reached by many configurations. At other frequencies, such as around 400 MHz, the chain covers a much broader range of voltages, with many finely spaced response levels. Such frequencies are especially valuable for computation, because they provide a large and densely populated output span that can be exploited to shape the weighted sum performed on RF inputs. More generally, the map shows that programming the binary magnetic state of the chain gives access to a broad reconfigurable family of transfer functions, including both positive and negative outputs and markedly different spectral profiles. Furthermore, the chain response is continuous in frequency over its whole operating range. It can therefore process multi-tone RF inputs as long as their Fourier components fall within this range, irrespective of the number of tones. This rich response landscape constitutes the computational resource that we exploit below for classification.

## Task-level reconfiguration of a remotely programmed RF neural network

We combine two 11-junction chains into a perceptron neural network with 22 synapses and two output neurons. As illustrated in Fig. 3a, the network is programmed chain by chain through an RF switch that selects which chain receives the write signal. During inference, RF input signals are sent through the strip line of each chain, in the same way as the programming signals. For each input waveform, classification is performed from the dc voltages $\Delta V_0$ and $\Delta V_1$ across the two chains after baseline subtraction using the response measured in a reference magnetic configuration (Methods). The predicted class is then given by the chain with the highest corrected voltage.

We use this platform to test whether the same hardware can be reassigned by remote RF programming to two qualitatively different classification tasks: distinguishing handwritten digits[31] ("0" versus "1") and identifying drones from their Wi-Fi RF signatures[32]. In both cases, the inputs are encoded as RF waveforms and synthesized by an arbitrary waveform generator before being injected into the network. For the digit task, each $8 \times 8$ image is converted into a 64-tone waveform whose amplitudes encode the pixel intensities. For the drone task, the inputs are reconstructed from measured RF spectra comprising 256 frequency bins (Methods). Signal amplitudes are chosen to avoid read-disturbance (Methods).

This encoding allows high-dimensional inputs to be processed through the same compact RF interface. Rather than assigning one physical input line to each pixel or spectral bin, as in conventional crossbar-style implementations, the information is multiplexed in frequency and processed by the continuous spectral responses of the synaptic chains, as shown in [17–23]. As a result, each chain of 11 synapses is able to process the 64 pixels images as well as the 256 bins spectra. In this way, a compact 11-by-2 network with binary programmable states provides a rich and continuously reconfigurable mapping of the input spectrum.

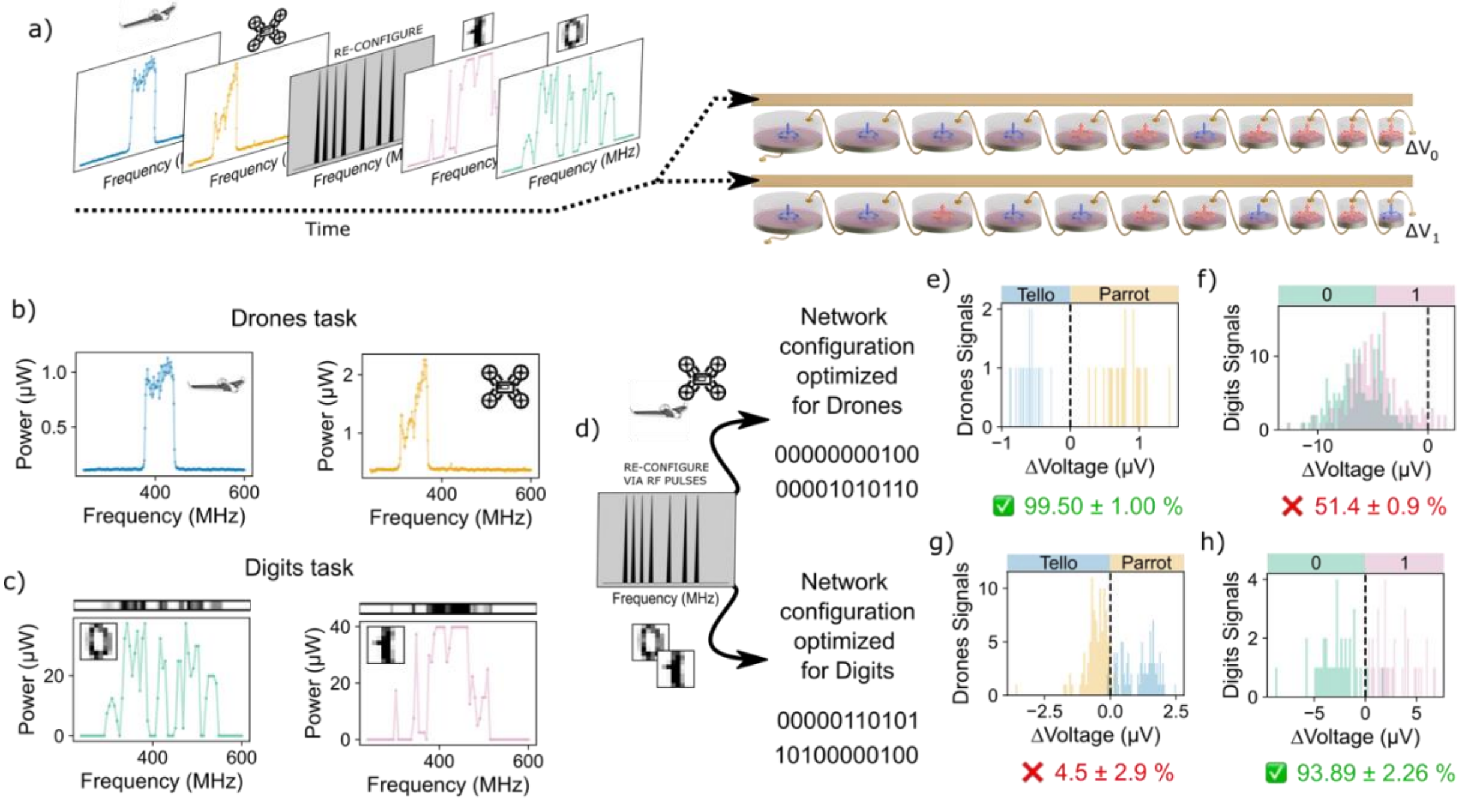

***Figure 3.*** *(a) Schematic of RF inputs and programming signals sent to the network through the strip line of each chain. Programming is performed chain-by-chain using an RF switch. (b) Examples of unrolled images of the digits dataset and corresponding spectra. (c) Examples of spectra of the drones signatures. (d) Schematic of programming pulses to reconfigure the network. (e-f-g-h) Histograms of the voltage difference between the response of chains 0 and 1, for the drones (e-g) and digits (f-h) tasks on different network configurations: (e-f) optimized for the drones task and (g-h) optimized for the digits task. The dashed vertical line corresponds to a difference of 0, which separates the histograms when classification is perfect. For e-h the histograms correspond to the test examples on the best fold while for f-g they correspond to the*

*whole dataset. The accuracies are indicated as averages and standard deviations over the five folds.*

In order to evaluate the inference capability of the spintronic network, we measure the response of the network in each possible configuration and for each example of the dataset (Methods). We perform an exhaustive search of the best configuration, i.e. with the lowest cross-entropy loss on the training set. Then we evaluate accuracy, i.e. fraction of correctly classified examples, on the test set. Figure 3e-f summarizes the two resulting optimal configurations. For each task, it shows the corresponding optimal configuration together with the distributions of $\Delta V = \Delta V_0 - \Delta V_1$, which reveal the separation between the two classes.

When programmed in the optimal configuration for drones classification, the network reaches 100.00 ± 0.00 % train accuracy and 99.50 ± 1.00 % test accuracy on the drones task (Methods), as shown in Figure 3e. The corresponding ΔV distributions for the "Tello" and "Parrot" classes are well separated on correct sides of the decision boundary, shown by the dashed vertical line. However, attempting classification of the digits task in this network configuration (Figure 3f) yields only 51.4 ± 0.9 % accuracy on the whole dataset. The distributions are not separated and the accuracy is at chance level.

We then re-program the network in the configuration optimal for handwritten-digit classification. As shown in Figure 3h, the network reaches 95.28 ± 0.56 % train accuracy and 93.89 ± 2.26 % test accuracy on the digit task (distributions well separated on the correct sides of the decision boundary). Its performance on the drones task (Figure 3g) drops to 4.5 ± 2.9 % accuracy on the whole dataset. The ΔV distributions are mostly separated, means that the network configuration that correctly classifies the digit task extracts features that are meaningful to the drone dataset. However, the ΔV distributions are on the incorrect sides of the decision boundary, leading to incorrect classification.

For each task, the optimal configuration performs similarly for the train and test sets, showing that the network does not overfit. Furthermore, the obtained test accuracies are comparable to those of a standard software binary perceptron: 95.56 ± 1.62 % on the digit task and 100.0 ± 0.0 % on the drone task (see Methods). The strong performance difference obtained after reprogramming for each task shows that remote RF writing does not merely fine-tune the network, but switches the same hardware between distinct functional regimes.

**In-situ training of the neural network**

To go beyond exhaustive characterization, which does not scale to large networks, we perform in-situ training. We employ an algorithm, proposed by Skalli et al. [33], which is compatible with binary training parameters and model-free, thus promising for future on-chip learning implementations. The core of the algorithm consists in flipping the polarity of randomly selected synapses. If the reconfiguration results in a lower training loss, the new configuration is preserved. Otherwise, the network is reverted in the previous configuration. The number of flipped synapses decreases with the loss (see Methods), mimicking annealing. As described in Figure 4a, the forward pass and programming are performed in hardware while the loss computation and comparisons are performed on a computer (see Methods).

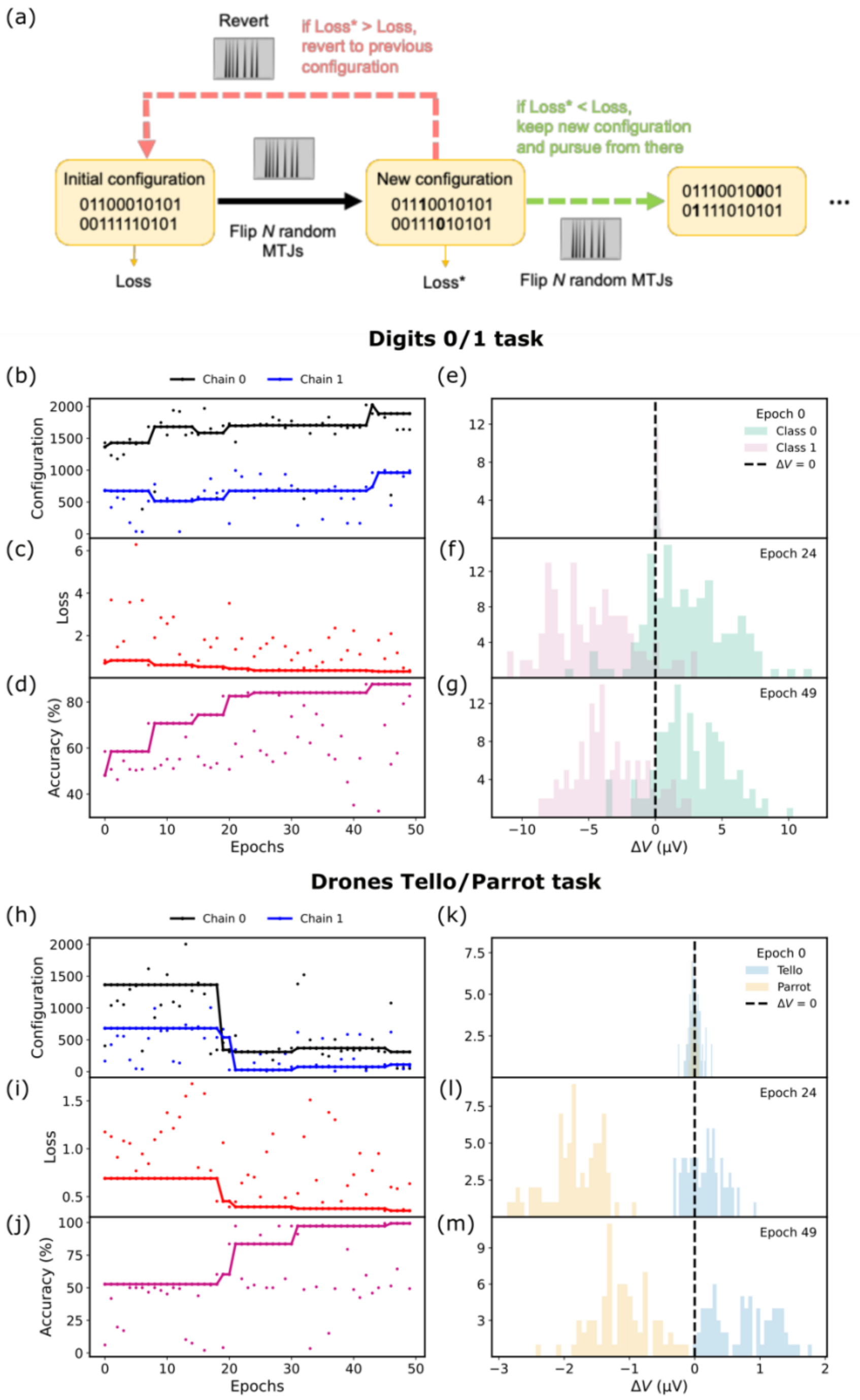

*Figure 4 (a) Schematic of the in-situ training process. (b, c, d) For the digits 0 versus 1 task, evolution through epochs of: (b) the configuration of each chain (colors) converted from the 11-bit binary code to an integer using the highest frequency junction as the least significant bit, (c) the training loss, (d) the training accuracy. The dots correspond to rejected configurations while the solid line corresponds to accepted configurations. (e, f, g) Histograms of the voltage difference between the response of chains 0 and 1 for the training examples at epoch 0 (e), epoch 24 (f) and epoch 49 (g). The dashed vertical line corresponds to a difference of 0, which separates the histograms when classification is perfect. Plots (h) to (m) replicate plots (b) to (g) for the drones Tello versus Parrot task.*

Figures 4b-c-d depict the chain configurations, training loss and training accuracy versus the number of epochs, for the digits task. The dots show the rejected configurations while the solid line shows the preserved configurations. Figures 4e-f-g show the distribution of output voltages for classes 0 and 1 at epochs 0, 24 and 49 respectively. After 50 epochs, we reach 87.88 % train accuracy and 90.00 % test accuracy. Using the same experimental in-situ training procedure (Figures 4h to 4m), the spintronic network reaches 100.0 % test accuracy on the "Tello-vs-Parrot" drone task. We perform two additional drone tasks: recognizing a "Tello" drone from a WiFi router with 99.14 % test accuracy and recognizing a "Spektrum DX4e" from a "Nineeg" drone with 94.99 % test accuracy (see Methods). The fact that the test accuracies are slightly above the train accuracies are chance effects due to the small numbers of examples. The accuracy of the spintronic network for each task is equivalent to a software binary perceptron trained with the standard BOP algorithm[34] (see Methods).

## Discussion

Compared with conventional in-memory approaches based on resistive or phase-change crossbars[1,2,5,8,10,35,36], our method removes individual electrical access from the programming path. Photonic and optoelectronic alternatives offer remote or parallel weight control through microring weight banks, all-optically controlled memristors, or spatial light modulators[4,11–16,37–39]. However, these approaches typically rely either on individual spectral calibration of resonators[37] or on spatial addressing with slow refresh rates, tight alignment and field-of-view constraints[39]. By contrast, the present platform combines non-volatile storage, broadcast electrical programming, frequency-selective addressing and electrical readout in the same device layer, using the same circuitry and shared access for both inference and programming. To our knowledge, this combination is presently unique among integrated neuromorphic hardware platforms.

This architecture can be scaled by increasing both the number of chains and the number of junctions within each chain. The latter is set mainly by the accessible frequency range and by the width of the individual switching windows. For magnetic vortices, analytical theory and experiments show that eigenfrequencies can be engineered from about 10 MHz to beyond 1 GHz depending on geometry and bias conditions[40]. In the present devices, the switching window is about several tens of MHz and arises from the effective magnetic field acting on the vortex, including both the applied field and internal inhomogeneities in the junction, such as stack imbalances and grain-induced disorder. With further optimization of materials and geometries so that the magnetic damping is about 1 to 2 %, the windows would be narrow enough to accommodate about 50 to 100 independently switchable devices per chain (Methods), and series connection of multiple chains can extend the total number of synapses to several hundreds. As the network size increases, reducing device dispersion will be key to maximizing computational power and training. The switching frequencies of each junction can be calibrated once entirely through the shared strip line, thus relaxing the requirements on fabrication uniformity.

Energy and speed can also be further improved. Here, programming is performed with continuous-wave microwave signals and sequential switching of the junctions. Resonant nanosecond pulses are an attractive route for scaled programming. By optimizing materials and field line geometry, Pigeau et al. switch a vortex core (dot diameter 500 nm, frequency close to 250 MHz) back and forth with −11 dBm pulses of 50 ns. This corresponds to a switching energy of the order of 5 pJ[28]. We have achieved vortex switching in MTJs with 1 dBm pulses of 25 ns, reaching 31 pJ switching energy (Methods). These numbers show that vortex-based RF MTJs are competitive with other in-memory computing technologies (Methods). Experiments and analytical simulations indicate that the switching energy scales linearly with the vortex frequency (Methods), giving guidelines for optimizing both diameters and strip lines[41]. Finally, with better engineering of the switching bands to suppress overlaps, several junctions could also be written in parallel rather than one after another.

The binary character of the present synapses is also not a fundamental limit. Elliptical nanomagnets can host metastable double-vortex configurations, and different double-vortex states exhibit distinct resonance modes[42,43]. This points to a natural route toward multistate resonant synapses based on richer non-uniform magnetic textures. At the other end of the design space, if footprint becomes the dominant constraints, the same resonant-switching concept could be transposed to sub-100-nm perpendicular junctions with more uniform magnetic states. Microwave-assisted switching has already been demonstrated in perpendicular nanodots, including layer-selective switching in coupled perpendicular structures[44].

The concept is also not restricted to tunnel junctions. Replacing MTJs by all-metallic stacks with GMR readout would open a route toward three-dimensional implementations based on electrodeposited multilayer nanowires[45,46]. Stack engineering makes is possible to remove the need for an external magnetic field[47]. Integrating the strip lines, as done in field-assisted MRAMs[48], will increase compacity. Equipping the chip with a broadband antenna, whether discrete or integrated such as shown in [49] would enable over-the-air network reconfiguration. Finally, both training networks with binary parameters[50] and training physical neural networks[51] are active research topics. Resulting algorithms, capable of training large-scale binary radiofrequency spintronic networks, will make it possible to tackle complex tasks. In that perspective, the most compelling application space for this architecture are native spectral tasks, such as wireless-signal identification, spectrum sensing, communication monitoring and edge RF intelligence.

**Conclusion**

This work establishes a neuromorphic architecture in which non-volatile binary magnetic states remotely reconfigure continuous spectral transfer functions. Using a single shared RF access for synaptic programming, state readout and RF input delivery, it removes the need for per-synapse access lines and selector devices and opens the route towards shared RF circuitry for programming and inference. This combination of compactness, scalability and functional reconfigurability makes resonant spintronic networks a promising route towards dense three-dimensional processors for RF-native edge intelligence.

# Acknowledgments

The authors thank A.F. Vincent and Damien Querlioz for their insightful comments. This work was supported by the European Union's Horizon 2020 Research and Innovation Programme under Grant RadioSpin No. 101017098. The text of the article was partially edited by a large language model (OpenAI ChatGPT).

# Methods

## Magnetic tunnel junction sample preparation

The magnetic tunnel junction (MTJ) stack composition is:
*Buffer [[5 Ta / 25 CuN]x6 / 5 Ta / 5 Ru] / Antiferromagnet [6 IrMn] / Synthetic Antiferromagnet [2 CoFe30 / 0.825 Ru / 2.6 CoFe40B20] / Tunnel barrier [MgO [15 Ohm um2]]/ Free layer [2.0 CoFe40B20 / 0.21 Ta / 40 nm CoFeSiB] / 10 Ta / 7 Ru (with the thickness in nm).*

The MTJ stack is deposited by magnetron sputtering using the Singulus Timaris MTM tool. After electron-beam lithography with a negative AR-N 7520.18 resist, the pillars are defined by ion beam milling. The Tunnel Magneto-Resistance is about 184 %.

The free layer is a composite with the following roles:
- CoFeB is necessary for having good MgO texture and subsequent large TMR values.
- Ta separates the crystallinity of the Co-based layers.
- Amorphous CoFeSiB reduces the impact of the defects on the vortex core dynamics[25].

**Chains.** We use two chains of eleven serially-connected synaptic junctions, where diameters vary from 300 nm to 800 nm by step of 50 nm. The MTJs are connected head-to-tail by optical lithography, as see in Figure 1c.

**Strip line for RF transmission.** The strip line is a 500 nm thick and 3µm wide AlSiCu strip line deposited on a layer of AlOx above the free layer of the MTJs. For the chains of 11 MTJs (Fig. 1f, Fig. 2, Fig. 3 and Fig. 4) the AlOx layer is 600 nm thick, leading to an efficiency of 0.66 Oe/mA. For the single device MTJ (Fig. 1d-e and Fig. S10), the AlOx layer is 50 nm thick, leading to an efficiency of 1.8 Oe/mA.

The resistance of the strip line is 7 ohms, and a 50 ohms resistance is connected in series with the field line to improve the matching with the 50 ohms impedance instruments and cables.
**Connections to the setup.** The contact pads of the chains and strip lines are connected to SMA ports by aluminium wire-bonding. The RF signals are injected into the strip lines through their pads.

## Experimental setup for the RF classification.

The complete experimental setup, described in *Figure S 1*, consists of two main blocks: the Arbitrary Waveform Generator (AWG) block in order to generate the signals for classification and the rest of the setup for programming and measuring the spintronic network.

**AWG block.** The AWG block is composed of an arbitrary waveform generator a protection circuit and an RF amplifier. The AWG output is first divided using a power splitter. Each output is sent to a circulator covering 240–400 MHz and 400–600 MHz, respectively. The two signals are then recombined using a second power splitter. The combined output passes through a 10 dBm attenuator followed by a +40 dBm RF amplifier.

A directional coupler is connected at the amplifier output: the main port provides the AWG block output, while the coupled port is connected to a power spectrum analyzer for monitoring. The RF amplifier is powered by a 12 V DC source. The attenuator minimizes reflections from the amplifier and prevents the formation of an RF cavity between the amplifier and the circulators. The circulators protect the AWG from reflections originating from the amplifier by allowing RF signals to propagate in only one direction and blocking the return path.

**Complete setup.** The output of the AWG block is connected to port 2 of RF switch A, while a separate RF signal generator is connected to port 1. The common (COM) port of switch A is connected to the COM port of switch B. Port 1 of switch B feeds the strip line of the first MTJ chain, and port 2 feeds the strip line of the second MTJ chain.
For electrical measurements, channel 1 of the DC current source and a nanovoltmeter are connected in parallel to the DC port of a bias-tee using a T-connector. The RF+DC port of the bias-tee is connected to the first MTJ chain via an SMA cable, while a 50 Ω load is connected to the RF port.
The same configuration is used for the second MTJ chain, employing a second nanovoltmeter, a second bias-tee, and channel 2 of the DC current source.

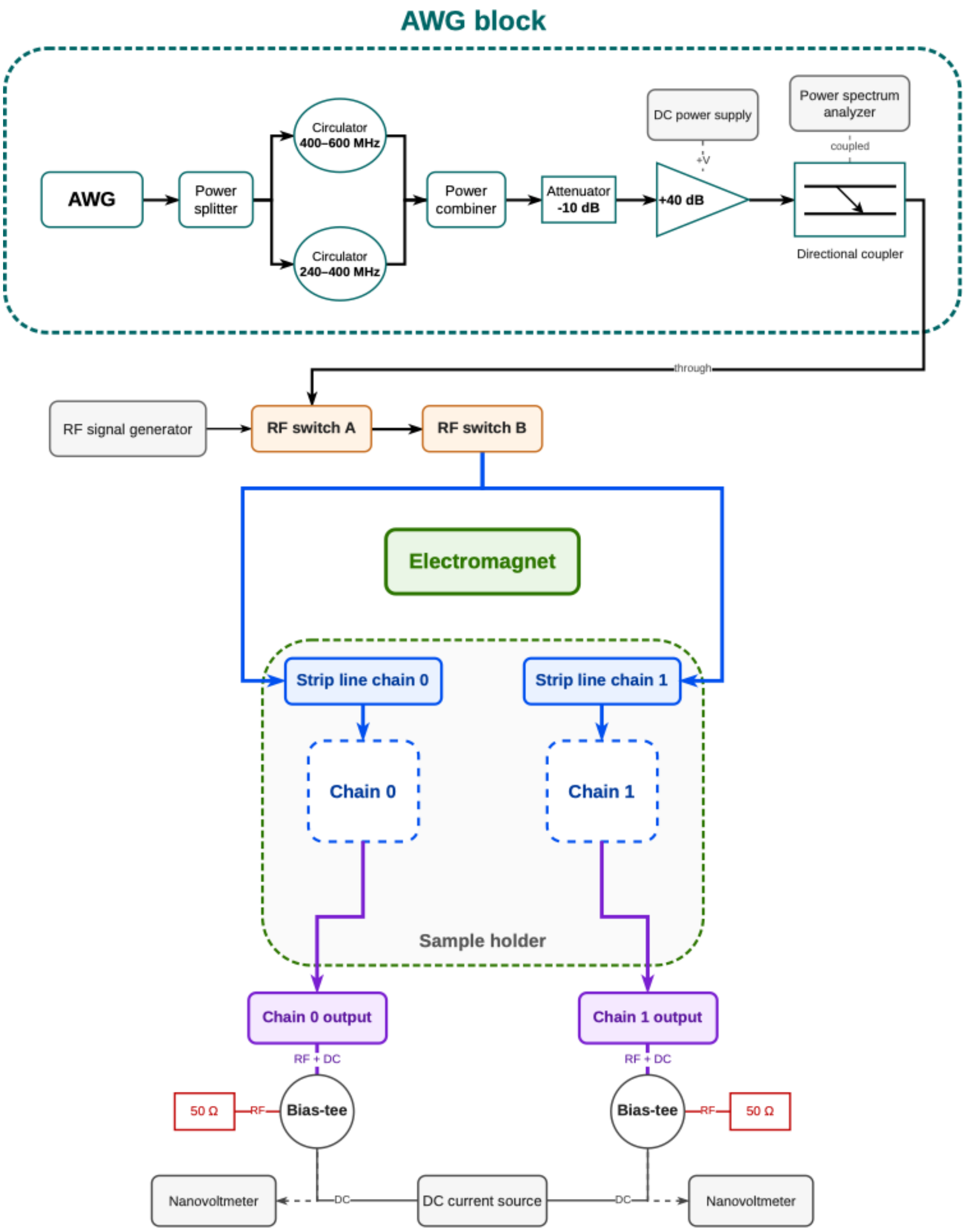

*Figure S 1 : Schematic of the experimental setup.*

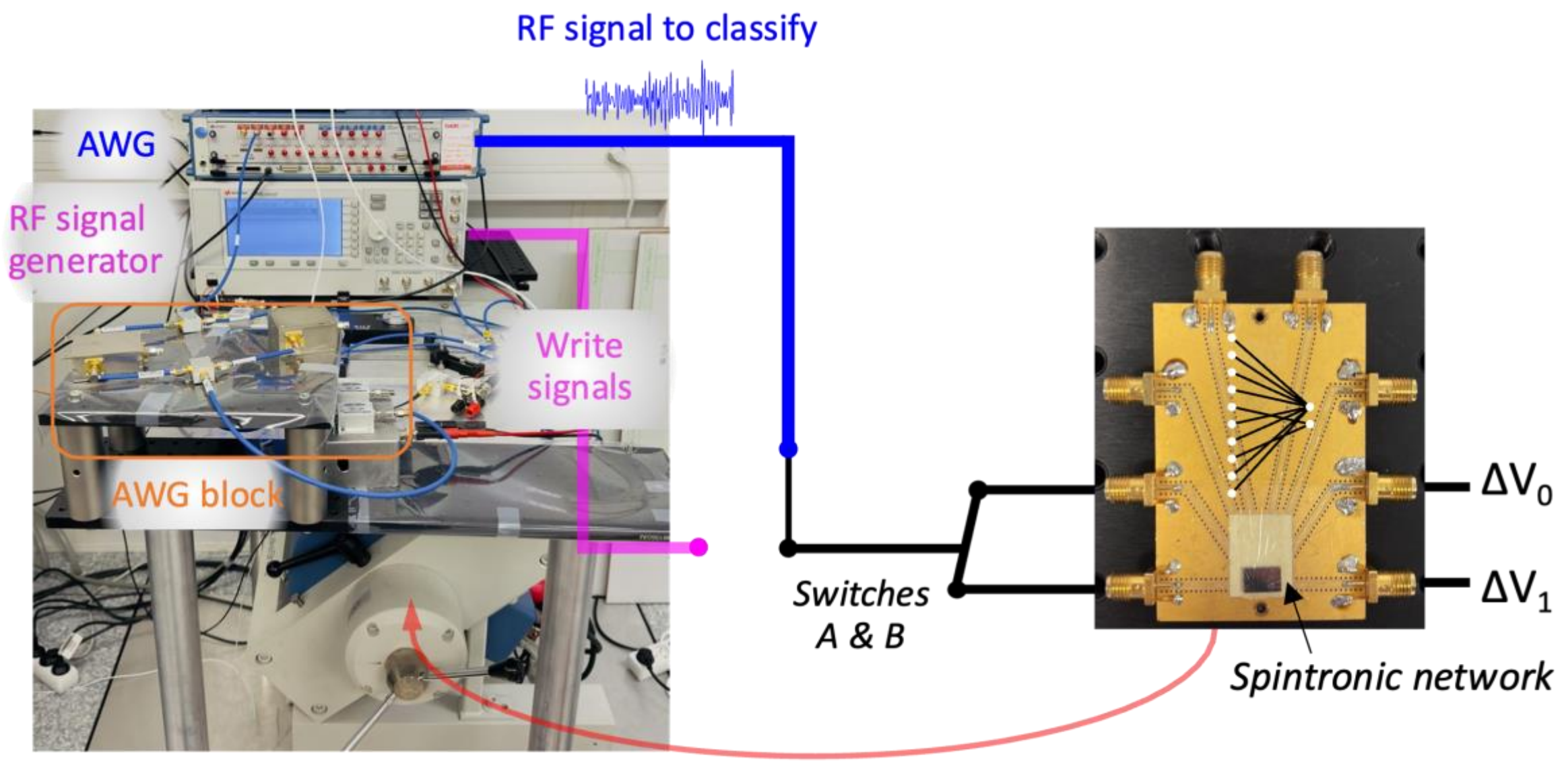

*Figure S 2: photograph of the experimental setup (nanovoltmeters and DC current source not shown).*

## Characterization of the magnetic tunnel junction chains

**Power dissipation correction.**

Wire-bonding leads to a frequency-dependent power loss in the strip line. We apply a RF power correction to the nominal inputs to correct for this loss, measured through the transmission factor $S_{21}$.

**Measurement of $S_{21}$ parameter.** The input and output of the strip line are connected to a Vector Network Analyzer (VNA) via SMA ports. The VNA compares the emitted signal, which goes through the wire-bonding and the antenna, to the received signal to estimate the transmission factor. Each chain's strip line has its own transmission factor, as shown in *Figure S 3*.

For the characterization of the chains, each power is corrected according to the $S_{21}$ of the measured chain. However, for signal classification, the dataset generation (Digits 0 and 1, Drones) is done with the mean value (in dB) of the two $S_{21}$.

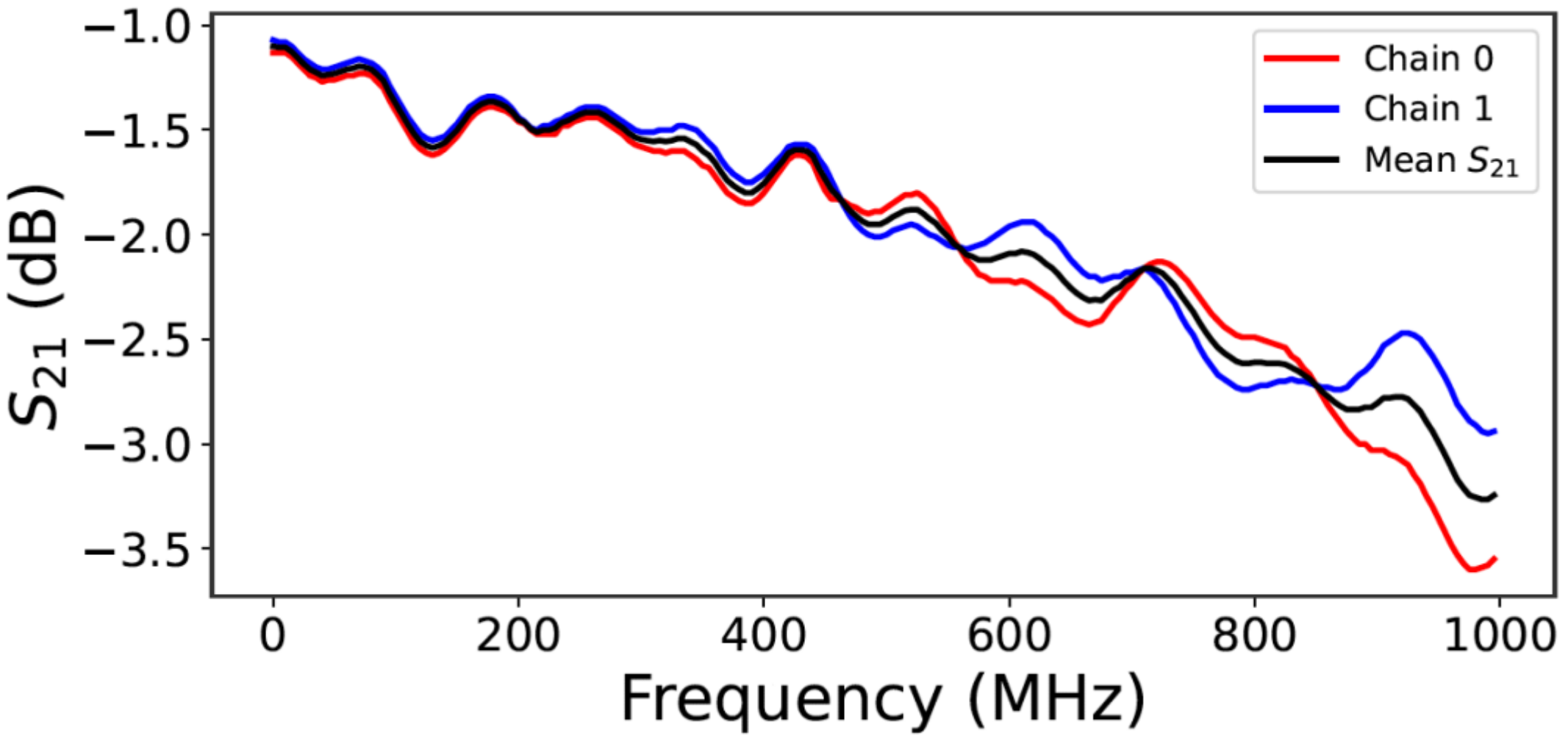

*Figure S 3: Transmission factor S21 versus frequency for chain 0 (red), chain 1 (blue) and their average (black).*

**Magnetic field characterization.** Circulators in the setup limit the usable frequency range to 240-600 MHz. Therefore, an out-of-plane magnetic field, perpendicular to the magnetic layers of the MTJs, is applied through an electro-magnet to shift the resonance frequencies into this range. In *Figure S 4*, we show the responses of the two chains to a sweep of magnetic field and frequency-swept RF signal. A field of -20mT satisfies the frequency limit condition for both chains.

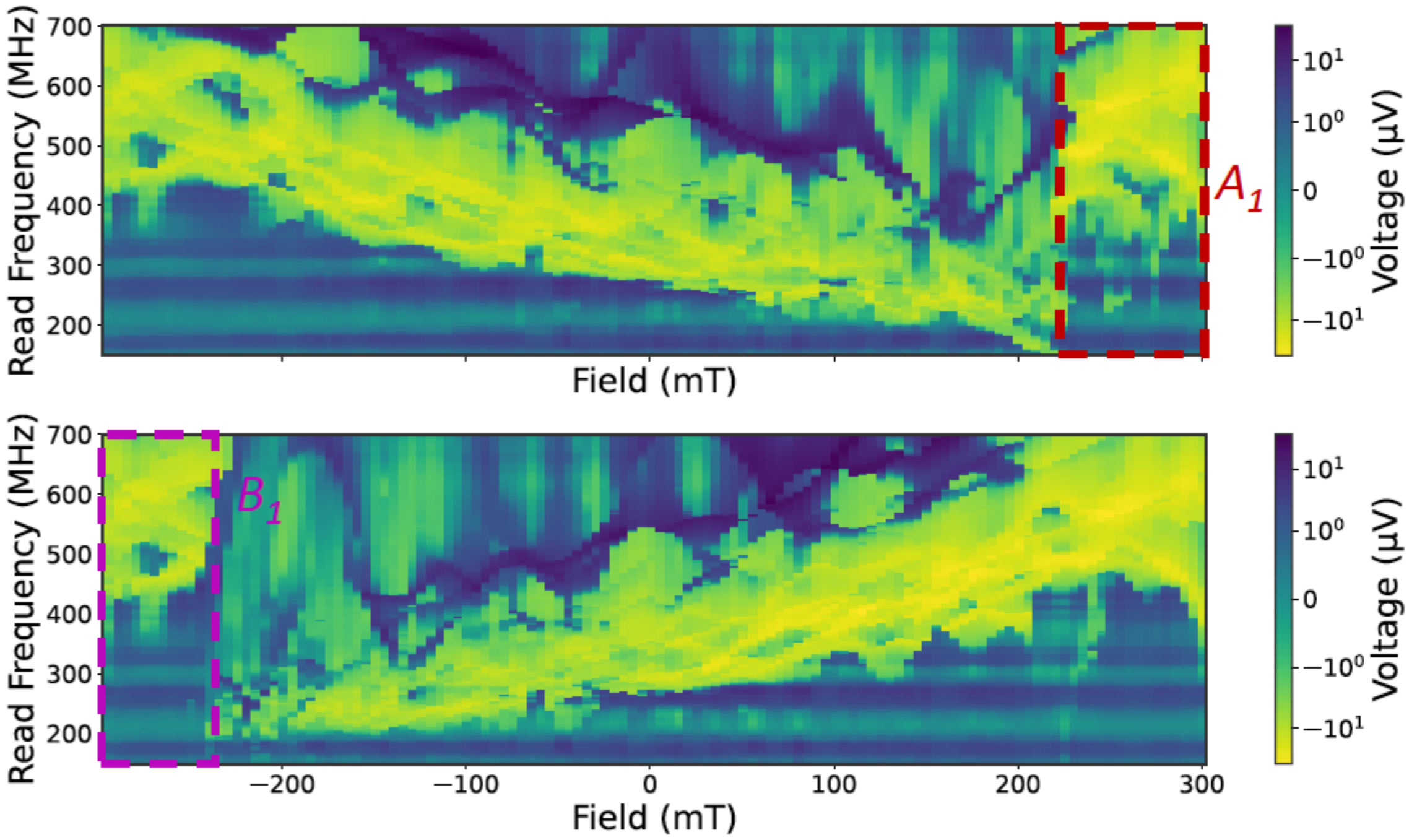

(a) Chain 0

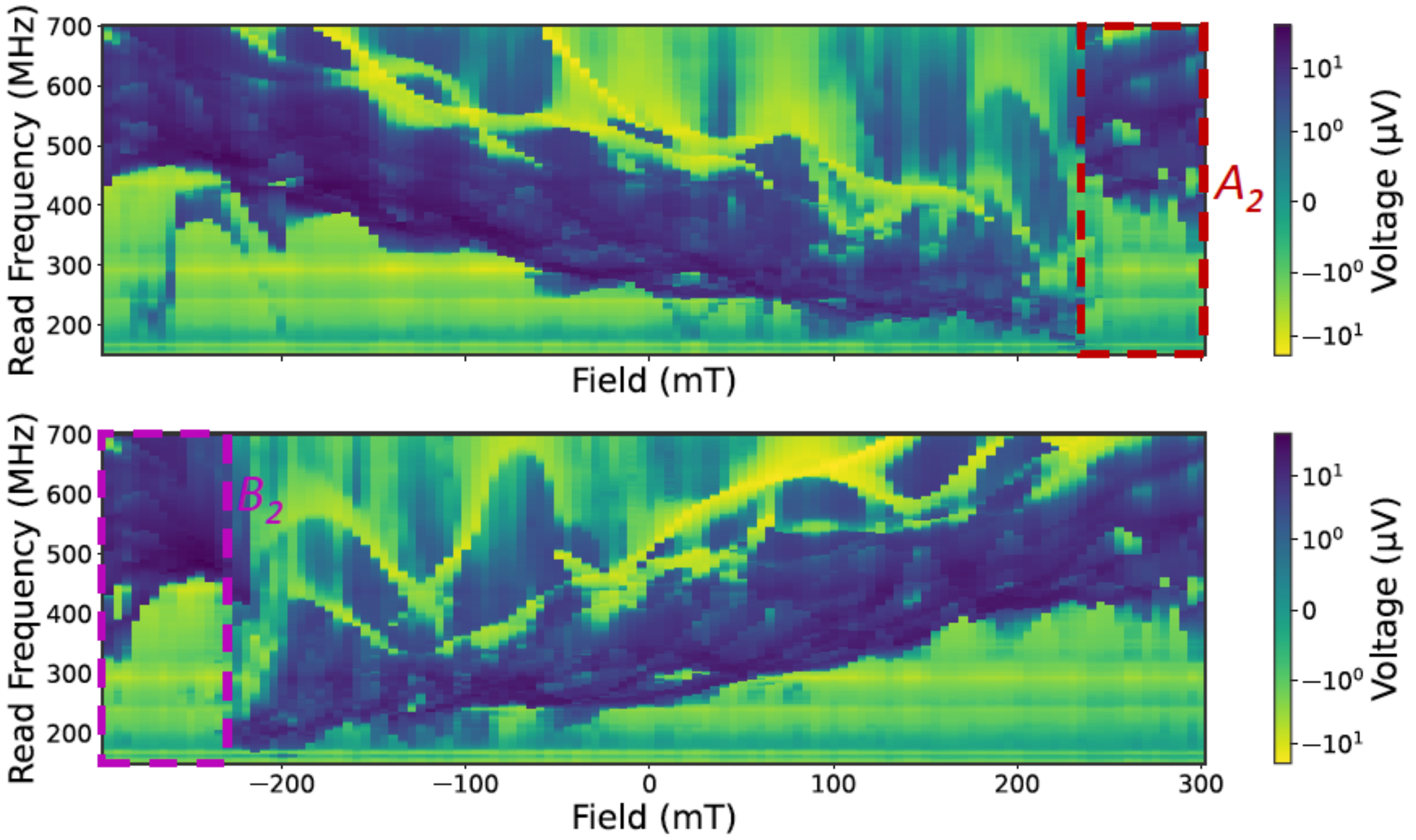

(b) Chain 1

*Figure S 4: Rectified spin-diode voltage responses for a perpendicular magnetic field swept between -300mT and +300 mT for (a) chain 0 and (b) chain 1. The field is first swept from -300mT to +300mT (top plots) and then from +300mT to -300mT (bottom plots).*
*A1 and B1 represent the regions where chain 0 configuration is fully-up or fully-down (respectively).*
*A2 and B2represent the regions where chain 1 configuration is fully-up or fully-down (respectively).*

**Determination of programming frequencies and powers.**
The chains are composed of synaptic devices of different diameters. The diameter defines the resonance frequency of the device. Thus, each device of the chain has a different resonance frequency. To reverse the vortex core on each device, the power and frequency to apply must be determined. These parameters can be predicted through micro-magnetic simulations, but the intrinsic dispersion of our devices requires experimental determination. Each device has a different sensitivity, and core reversal occurs at different powers and frequencies.
The procedure below describes the obtention of the map of Figure 1f.
We attempt programming at frequencies from 240 MHz to 600 MHz with a 1 MHz step. For each programming frequency, we:

1) Reset the chain in the fully down state with a – 300 mT perpendicular magnetic field.
2) Apply a 0.5 s RF pulse at the selected frequency.
3) Measure the spin-diode voltage response of the chain for the 240 MHz to 600 MHz range, at the reading power of -8 dBm.

This procedure provides a map of the spin-diode response versus programming and reading frequencies. We repeat the procedure for programming powers of 2, 5, 7 and 10 dBm, as shown in Figure S 5(a, b, c, d).
We then stitch the maps obtained at different powers into a single map, as shown in Figure S 5(e) and displayed in Figure 1f, which provides distinct frequencies to program each MTJ from the down state to the up state.

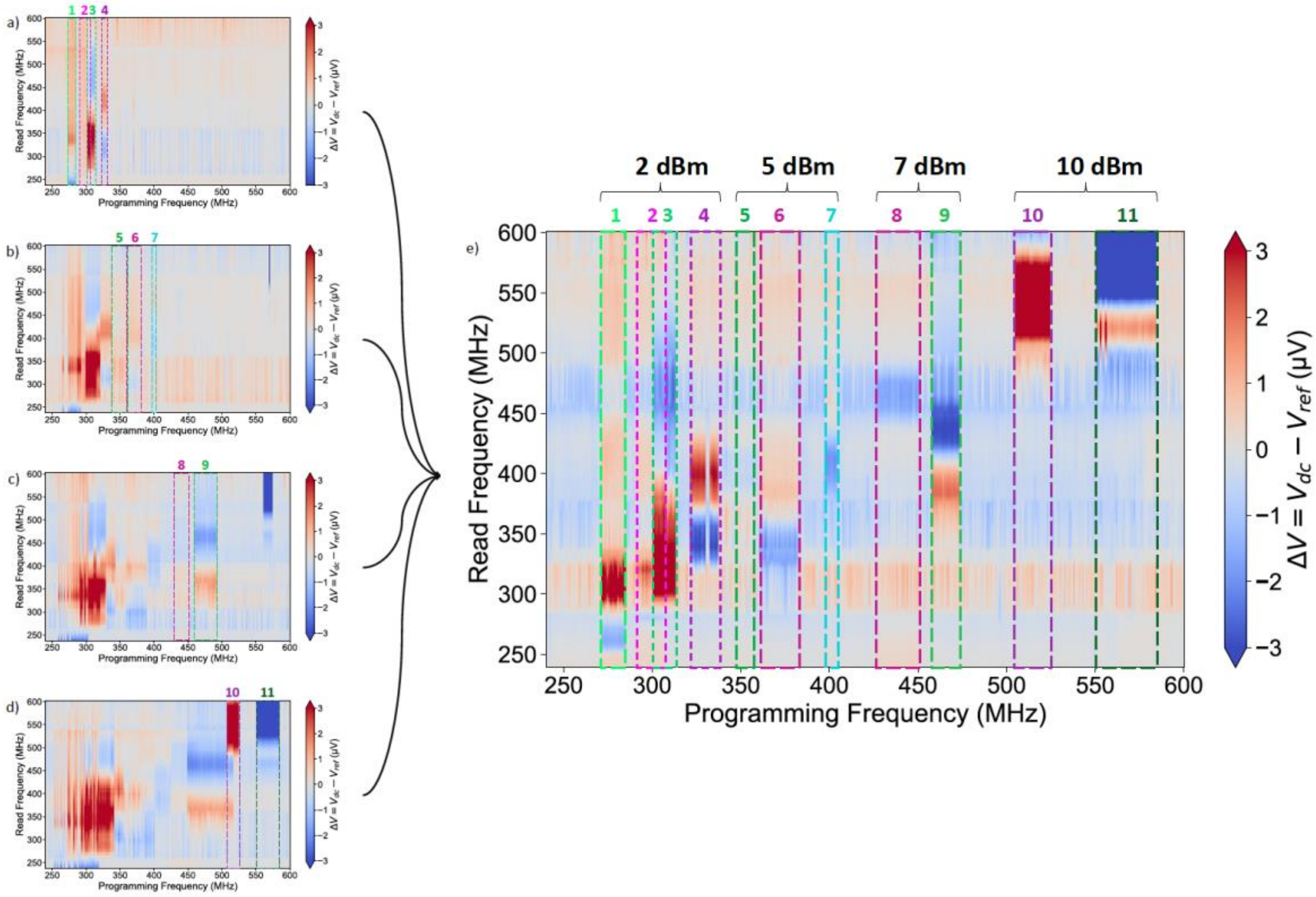

*Figure S 5: Response of the chain ΔV (voltage difference compared to the reference fully-down state) versus programming frequency and read frequency. The experiment is performed under a -20 mT perpendicular magnetic field. The color scale indicates the voltage difference ΔV between the chain response after the write attempt and that of the reference state. The respective switching zones for each of the 11 junctions are highlighted through dotted lines of different colors. The maps were measured using programming powers of (a) 2 dBm, (b) 5 dBm, (c) 7 dBm, and (d) 10 dBm. (e) Combined programming map constructed from the four measurements.*

We repeat this process starting from the fully up state in order to obtain the powers and frequencies to program each MTJ from the up state to the down state, resulting in Table 1 below. In order to evaluate the reliability of programming, we attempt programming 10 times for each {power, frequency} couple. We obtain 100 % success for each couple.

*Table 1: Programming powers and frequencies of chain 0.*

| MTJ | Frequency up to down (MHz) | Frequency down to up (MHz) | Power (dBm) |
|---|---|---|---|
| 1 | 260 | 275 | 2 |
| 2 | 278 | 298 | 2 |
| 3 | 292 | 311 | 2 |
| 4 | 305 | 325 | 2 |
| 5 | 327 | 347 | 5 |
| 6 | 346 | 365 | 5 |
| 7 | 376 | 400 | 5 |
| 8 | 410 | 442 | 7 |
| 9 | 433 | 465 | 7 |
| 10 | 485 | 512 | 10 |
| 11 | 522 | 560 | 10 |

We perform the same determination procedure for the second chain and obtain the programming {power, frequency} couples summarized in Table 2. MTJ 8 has a defect and cannot be switched.

*Table 2: Programming powers and frequencies of chain 1.*

| MTJ | Frequency up to down (MHz) | Frequency down to up (MHz) | Power (dBm) |
|---|---|---|---|
| 1 | 250 | 276 | 2 |
| 2 | 263 | 280 | 2 |
| 3 | 285 | 303 | 3 |
| 4 | 295 | 317 | 3 |
| 5 | 322 | 342 | 5 |
| 6 | 333 | 365 | 5 |
| 7 | 355 | 388 | 5 (up to down) /7 (down to up) |
| 8 | – | – | – |
| 9 | 422 | 450 | 10 |
| 10 | 467 | 500 | 10 |
| 11 | 522 | 560 | 10 (up to down) /13 (down to up) |

Once this determination is complete, programming is done exclusively by RF pulses, with a constant perpendicular magnetic field of -20mT.

**Programming any configuration with RF pulses.**
The programming frequency regions of some MTJs overlap. In order to reliably program the chain into any chosen configuration, we use the following procedure:

1) We set the chain in the fully down configuration by applying a series of RF pulses, i.e. the {power, frequency} couples of Table 1 from highest to lowers frequency.
2) Following Table 1, we apply a series of RF pulses to program the required MTJs in the up state.

## RF classification

**Handwritten digits recognition.** The task is distinguishing between images of “0” and “1”. The 360 images of 8-by-8 pixels each are from the dataset “Digits”. We send the inputs to the network as RF waveforms as schematized in Fig 3a (as shown in similar Leroux et al. [20]). The waveform for each image is constructed by assigning a frequency to each pixel:

$$S(t) = \sum_{i=0}^{63} S_i \times \cos(2\pi \times f_i \times t + \varphi_i) \quad (Eq\ 1)$$
$$\text{With } S_i = A_i \times S_{21}(f_i)$$

The frequencies $f_i$ are linearly spaced over the operating range of the synaptic chains (240-600 MHz). $A_i$ is proportional to the normalized intensity of pixel i, with a scaling factor of -14 dBm. $S_{21}(f_i)$ is a corrective parameter corresponding to mean value of $S_{21}$ parameters of both chains. $\varphi_i$ is a random phase at frequency $f_i$. The waveform is synthetized by the arbitrary waveform generator and sent to each chain through its strip-line. The whole image is sent as a single waveform, which differs from crossbar arrays implementations where an image of 64 pixels would require 64 input lines.

**Drones RF signatures.** The task is distinguishing between two types of drones ("Tello" and "Parrot") from their RF signature. The 200 RF signatures are spectra with 256 frequency bins, adapted from Basak et al.[32]. As for the image classification task, for each we use the arbitrary waveform generator to synthetize each waveform according to Eq. 1 and set the frequencies $f_i$ between 240 and 600 MHz. Here $A_i$ corresponds to the spectrum power at the i-th frequency, with a scaling factor of -20 dBm.

**Preventing read-disturbance.** We choose the amplitude of the inference signals so that they do not accidently perturb the configuration of the MTJs. We calibrate the maximum power using the image with the largest cumulated power.

- For the digits task, the image with the largest cumulated power reaches Power = 0.33dBm, which is below the switching threshold in each chain.
- For the drone task, the image with the largest cumulated power reaches Power = - 5.76dBm, which is below the switching threshold in each chain.

**Classification read-out.** The two outputs of the network are the voltages across chains 0 and 1. For each chain we consider the voltage difference between the measured configuration and a reference state. The reference states for chains 0 and 1 are respectively 10101010101 and 01010101010. These states with alternating up and down MTJs were chosen to maximize the expressivity of the chains. The prediction of the network corresponds to the index of the chain with the highest voltage difference, i.e. one-hot encoding.

**Classification accuracy.** For each task, we split the dataset into a train set (75 % of signals) and a test set (remaining 25 %). The accuracy is given as the percentage of correctly classified signals. For the reconfiguration study, we perform 5-fold cross-validation: the accuracy is the average over five random train/test splits. The errors are the standard deviation over the five splits.

## In-situ training

### Annealing-inspired algorithm

1) Initialize the system in the reference state.
2) Compute the initial Cross Entropy loss $L_{old}$ over the whole training set.
3) For each epoch:
   Randomly select $N = round\left(f_{switch}(L_{old})\right)$ MJTs and flip their states.
   Compute the new loss $L_{new}$.
   **if** $L_{new}. < L_{old}$ **then** $L_{old} := L_{new}$.
   **else:** restore the previous configuration.

The number of randomly selected MTJs N depends on the loss:

$$f_{switch}(Loss) = \frac{c}{1 + exp(-k \times (Loss - L_0))}$$

With c = 5, k = 5 and $L_0$ = 0.03, empirically chosen parameters resulting in the function shown in Figure S 6.

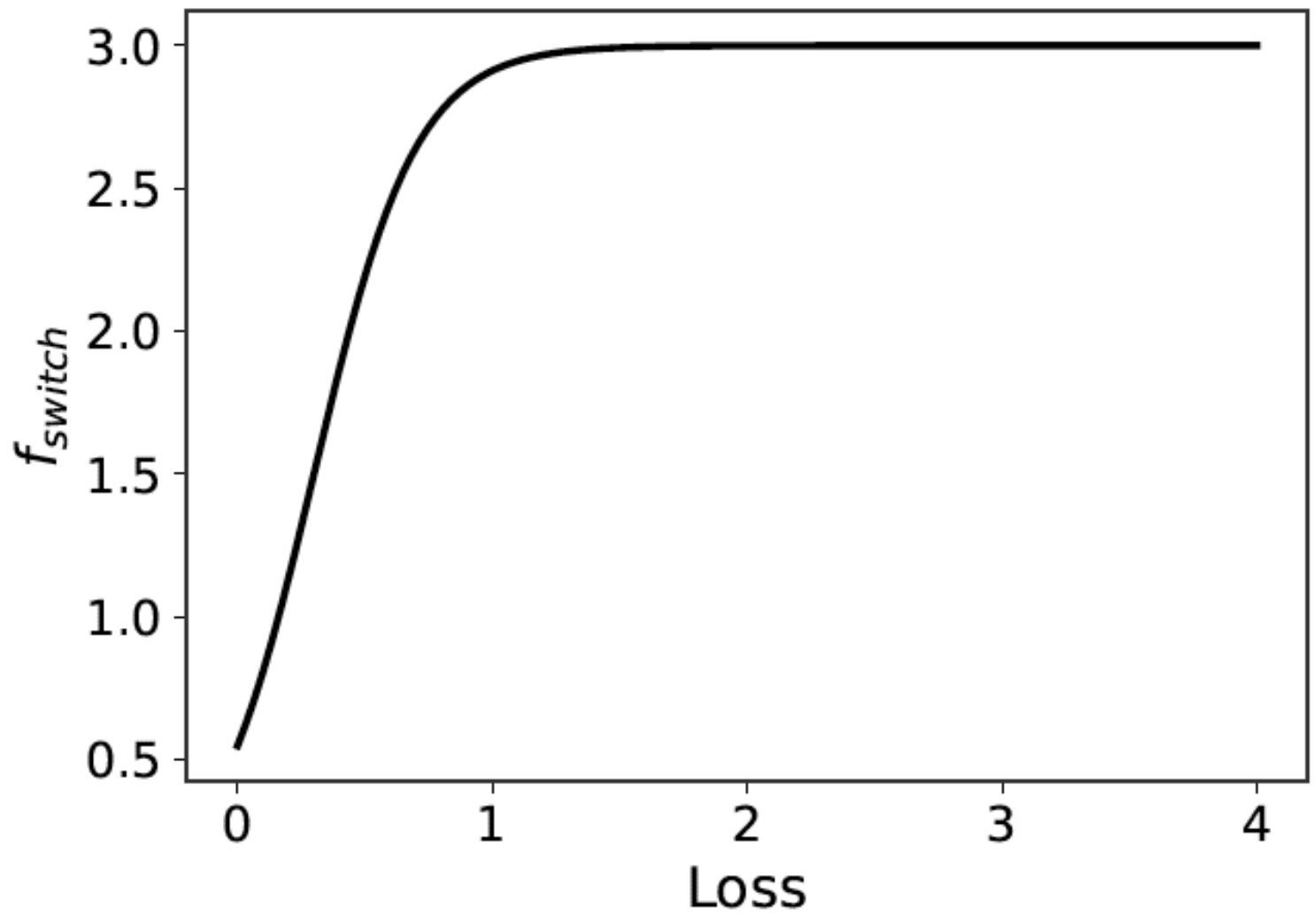

*Figure S 6 : Function to choose the number of MTJs to flip depending on the loss.*

**In-situ training of additional RF fingerprinting tasks**

We perform experimental in-situ training for two additional drone tasks. Like the Tello versus Parrot task, they consist in distinguishing two classes of emitters. The classes are all extracted from the Basak et al. dataset[32] and prepared as described in the ***RF classification*** section. For these tasks, we use the accuracy rather than the loss as criterion to keep or reject configurations.

We perform the following tasks:

- Tello drone versus WiFi router, reaching 99.14% test accuracy (Figure S 7).
- Spektrum DX4e drone versus Nineeg drone, reaching 94.99 % test accuracy (Figure S 8).

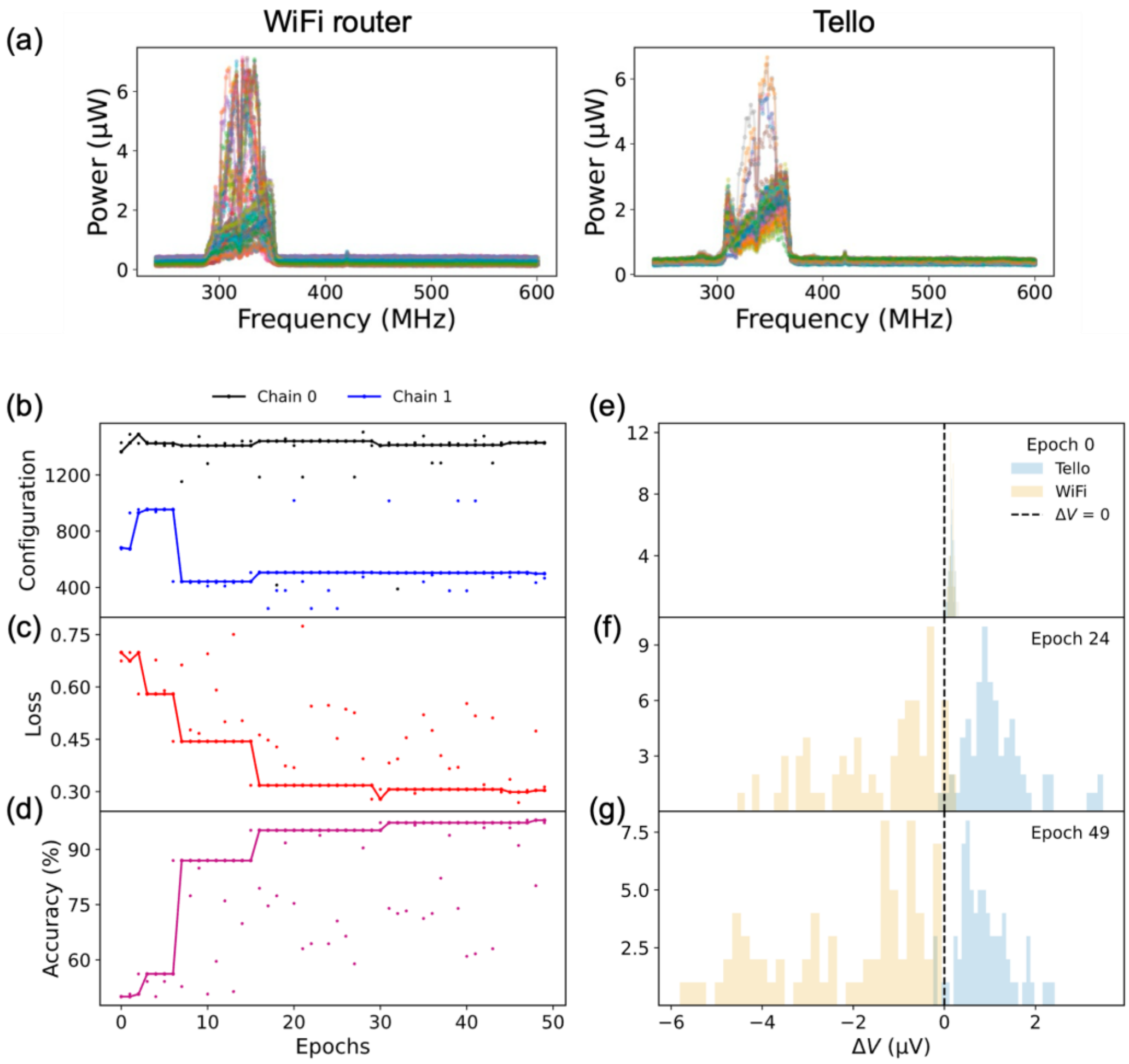

*Figure S 7 In-situ training of the Tello versus WiFi router task. (a) Spectra of the RF signals to classify. Each color is one example. (b, c, d) Evolution through epochs of: (b) the configuration of each chain (colors), (c) the training loss, (d) the training accuracy. The dots correspond to rejected configurations while the solid line corresponds to accepted configurations. (e, f, g) Histograms of the voltage difference between the response of chains 0 and 1 for the training examples at epoch 0 (e), epoch 24 (f) and epoch 49 (g). The dashed vertical line corresponds to a difference of 0, which separates the histograms when classification is perfect.*

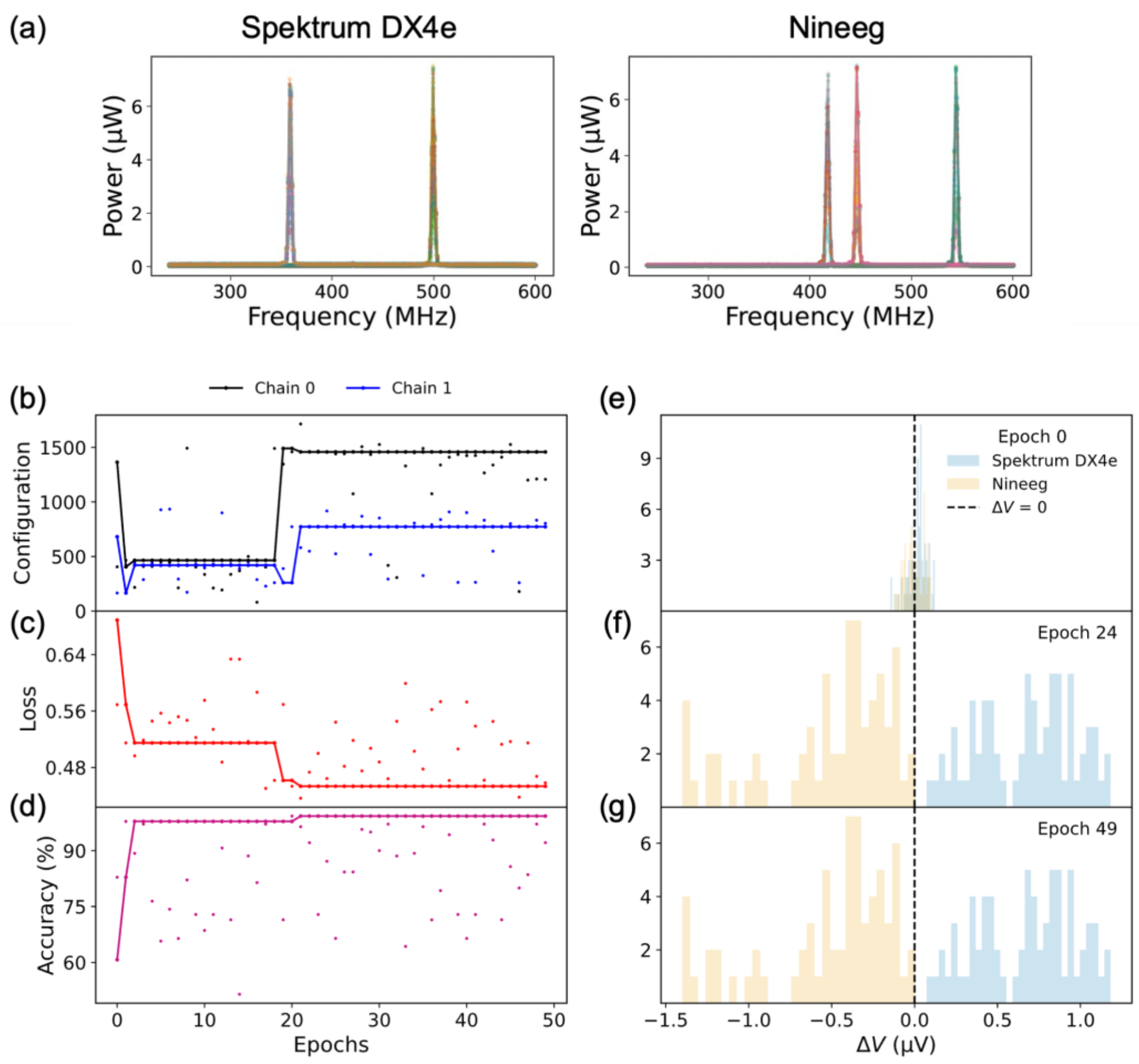

*Figure S 8 In-situ training of the Spektrum DX4e versus Nineeg task. (a) Spectra of the RF signals to classify. Each color is one example. (b, c, d) Evolution through epochs of: (b) the configuration of each chain (colors), (c) the training loss, (d) the training accuracy. The dots correspond to rejected configurations while the solid line corresponds to accepted configurations. (e, f, g) Histograms of the voltage difference between the response of chains 0 and 1 for the training examples at epoch 0 (e), epoch 24 (f) and epoch 49 (g). The dashed vertical line corresponds to a difference of 0, which separates the histograms when classification is perfect.*

## Comparison with a standard software network

The spintronic network here is a 11x2 single-layer network with binary parameters (vortex polarities) and no biases. For meaningful comparison, we chose as baseline the performance of a standard software 11x2 single-layer perceptron with binary (±1) weights and no biases.

In order to transform the inputs from dimension 64 (digits) and 256 (drones) to dimension 11, we can either perform block-binning or a sliding average. We evaluate both options.

**Inference**

Just as for the spintronic network, we perform an exhaustive search of the optimal configuration on the train set and evaluate the configuration on the test set. We use 5-fold cross-validation with a 75/25 train/test split. We obtain the following performance (average ± standard deviation over the five folds):

| | Software network | | Spintronic | |
|---|---|---|---|---|
| Task | Train accuracy | Test accuracy | Train accuracy | Test accuracy |
| Digits | Block: 93.40 ± 0.73 %<br>Sliding: 97.71 ± 0.28 % | Block: 92.50 ± 2.58 %<br>Sliding: 95.56 ± 1.62 % | 95.28 ± 0.56 % | 93.89 ± 2.26 % |
| Drones Tello vs Parrot | Block: 100.00 ± 0.00 %<br>Sliding: 100.00 ± 0.00 % | Block: 100.00 ± 0.00 %<br>Sliding: 100.00 ± 0.00 % | 100.00 ± 0.00 % | 99.50 ± 1.00 % |

We observe that the performance of the spintronic network is comparable to the best performance of the software network, as summarized in

Figure S *9*.

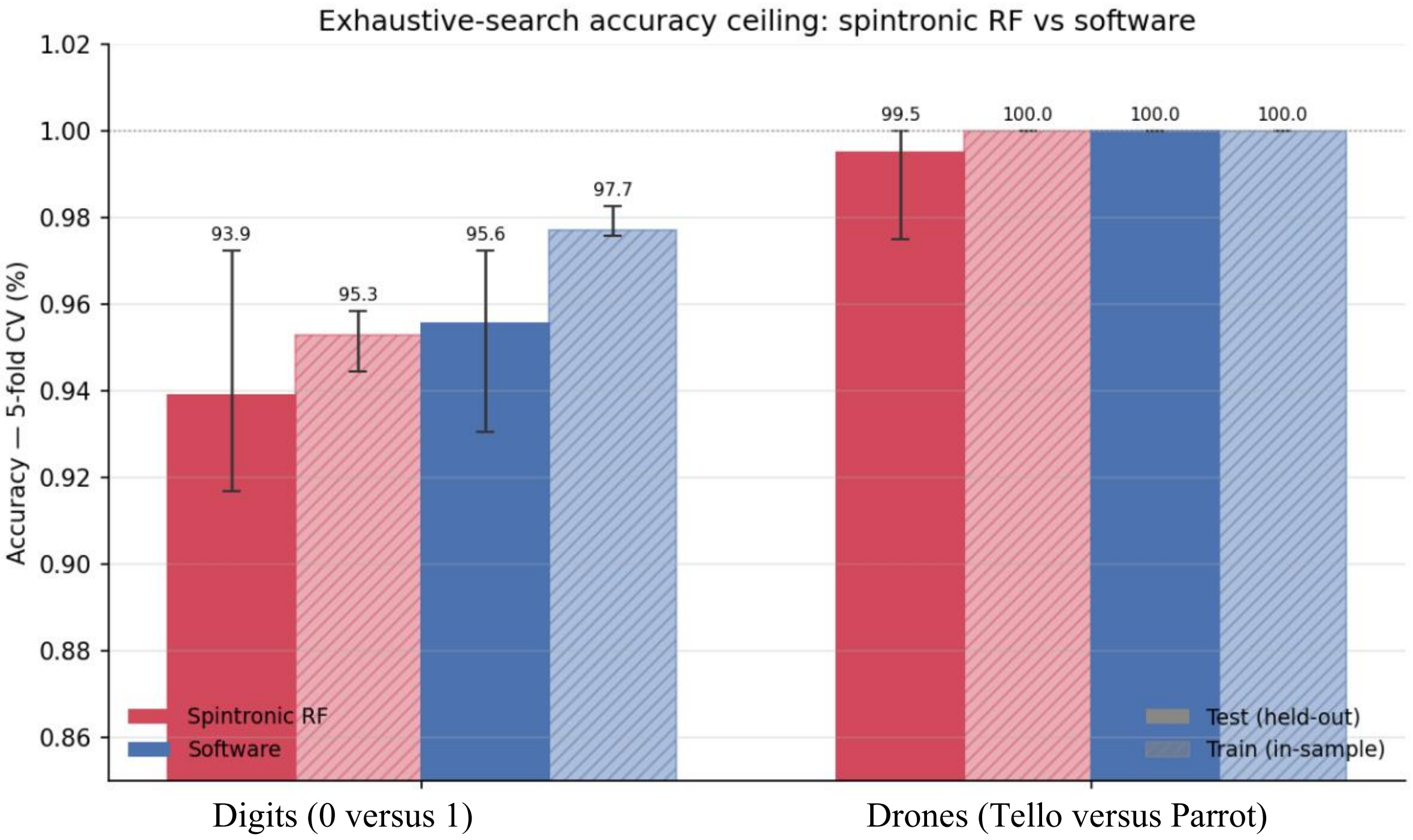

*Figure S 9 : Comparison of the spintronic RF network (red) and standard perceptron (blue) for each task. The test and train accuracies are respectively represented by solid and hashed bars. The bars represent the average over the five folds. The whiskers represent the min and max values over the five folds.*

## Training

For the software network, we use the BOP optimizer[34] which is state of the art to train binary neural networks. We use 5-fold cross-validation, with a 75/25 train/test split. We use Optuna[52] and 200 epochs to optimize the hyperparameters of BOP for each task.

The software and spintronic rf networks compare as follows:

| | Software network | Spintronic<br>(one run in-situ training) |
|---|---|---|

| Task | Train accuracy | Test accuracy | Train accuracy | Test accuracy |
|---|---|---|---|---|
| Digits | Block: 88.61 ± 1.13 %<br>Sliding: 89.86 ± 2.62 % | Block: 87.78 ± 7.58 %<br>Sliding: 90.00 ± 2.22 % | 87.88% | 90.00% |
| Drones Tello vs Parrot | Block: 98.38 ± 2.67 %<br>Sliding: 100.00 ± 0.00 % | Block: 100.00 ± 0.00 %<br>Sliding: 100.00 ± 0.00% | 99.3 % | 100 % |
| Drones Spektrum DX4e vs Nineeg | Block: 98.75 ± 1.31 %<br>Sliding: 83.75 ± 9.56 % | Block: 98.50 ± 3.00 %<br>Sliding: 87.00 ± 6.96 % | 99.29 % | 94.99 % |
| Drones Tello vs WiFi | Block: 80.88 ± 11.92 %<br>Sliding: 98.25 ± 1.21 % | Block: 80.50 ± 13.08 %<br>Sliding: 97.50 ± 2.74 % | 97.90 % | 99.14 % |

We conclude that the performance of the spintronic RF network is on-par with the standard software network.

## Programming with short RF pulses

We experimentally demonstrate the programming of individual devices with short RF pulses. Figure S 10 shows the probability of correctly programming the MTJ in each polarity versus the length of the RF pulse. We see that we can achieve reliable programming with RF pulses as short as 25 ns.

The programming power is 1 dBm. In consequence the programming energy is about 31 pJ.

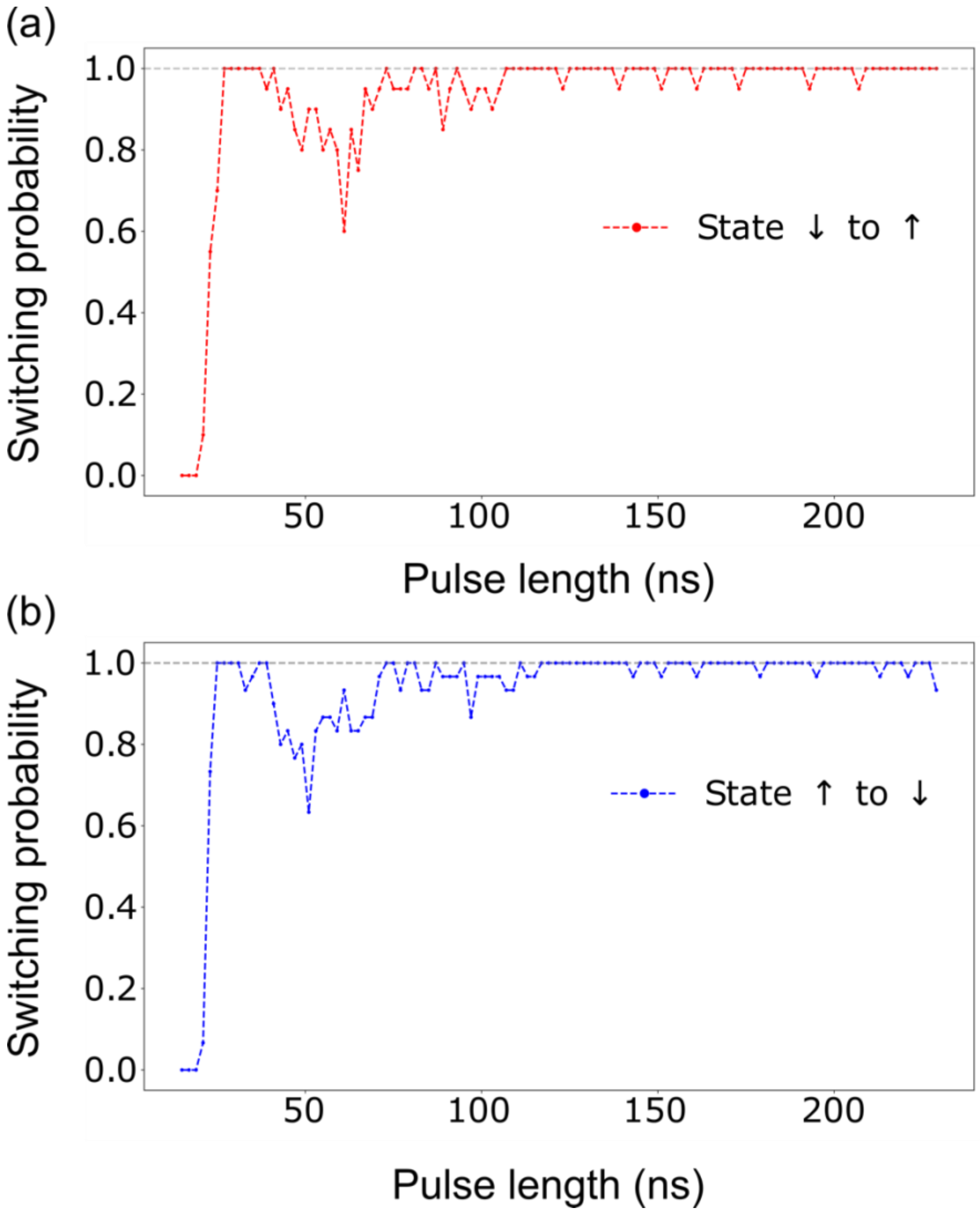

*Figure S 10. Probability to correctly program the MTJ in the positive (a) or negative (b) polarity, versus the pulse length. For both polarities the programming power is 1 dBm and the probability is computed over independent 20 trials. The programming frequencies are respectively 337 MHz and 358 MHz for the positive and negative polarities.*

## Programming power and energy as a function of vortex frequency and device diameter

To explore the dependence of the write conditions on junction frequency, we combine the analytical scaling of the vortex gyrotropic mode with the universal criterion for resonant core reversal. For thin circular dots, the gyrotropic angular frequency is well approximated by[53]:

$$\omega_G \simeq C \times \gamma \times M_s \times \frac{L}{R}$$

Where C is material dependent, $\gamma$ is the gyromagnetic ratio, Ms is the saturation magnetization, L is the thickness of the layer and R the radius of the dot. In consequence, at fixed material and thickness, the frequency $f_G$ is:

$$f_G \propto \omega_G \propto \frac{1}{R}$$

This relationship is consistent with the experimental data, represented in Figure S 11.

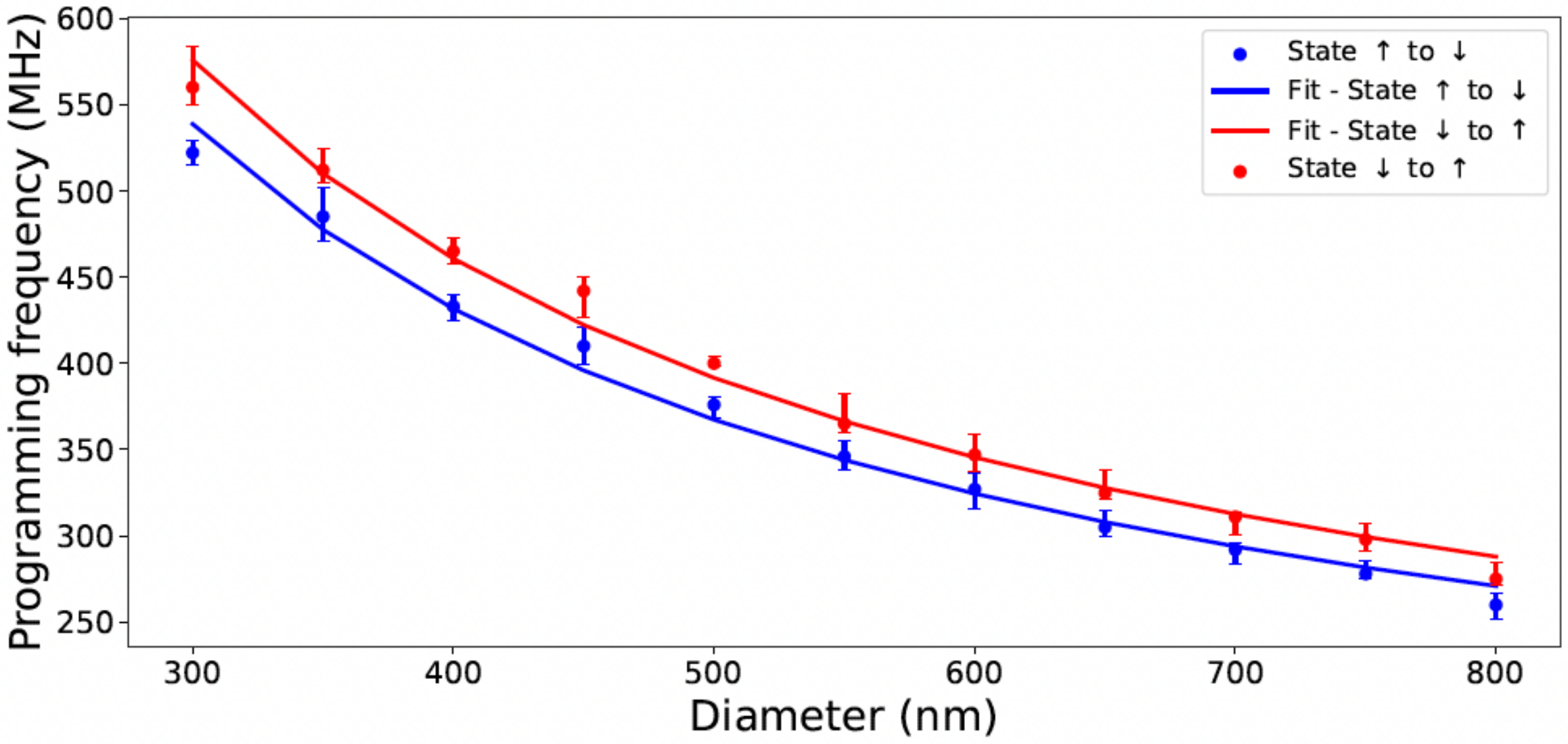

*Figure S 11 : Programming frequency for the two vortex-core transitions. Blue symbols correspond to the ↑ to ↓ switching and red symbols to the ↓ to ↑ switching. Whiskers corresponds to the switching regions of Figure 1f. The solid curves show the corresponding fits.*

Vortex-core reversal occurs when the core velocity reaches a critical value $v_c$, which is approximately independent of dot size. In the resonant limit, the minimum switching field obeys[54]:

$$H_{sw}^{min} \simeq \frac{d \times v_c}{3\gamma \times R}$$

with $d \simeq \alpha \left(1 + \frac{ln\left({}^{R}/_{R_c}\right)}{2}\right)$ where $\alpha$ is the damping parameter and $R_c$ is the vortex-core radius. Neglecting the logarithmic term, this gives:

$$H_{sw}^{min} \propto \frac{1}{R}$$

Assuming that the microwave field generated by the strip line is proportional to the RF current amplitude, the threshold write power then scales with the square of the frequency:

$$P_{sw} \propto \left(I_{sw}^{min}\right)^2 \propto \left(H_{sw}^{min}\right)^2 \propto (f_G)^2$$

This relationship is consistent with the experimental data, represented in Figure S 12. Our goal when determining programming powers was to have a small set of different power values rather than to find the minimum programming power for each MTJ. In consequence, the experimental programming powers form steps.

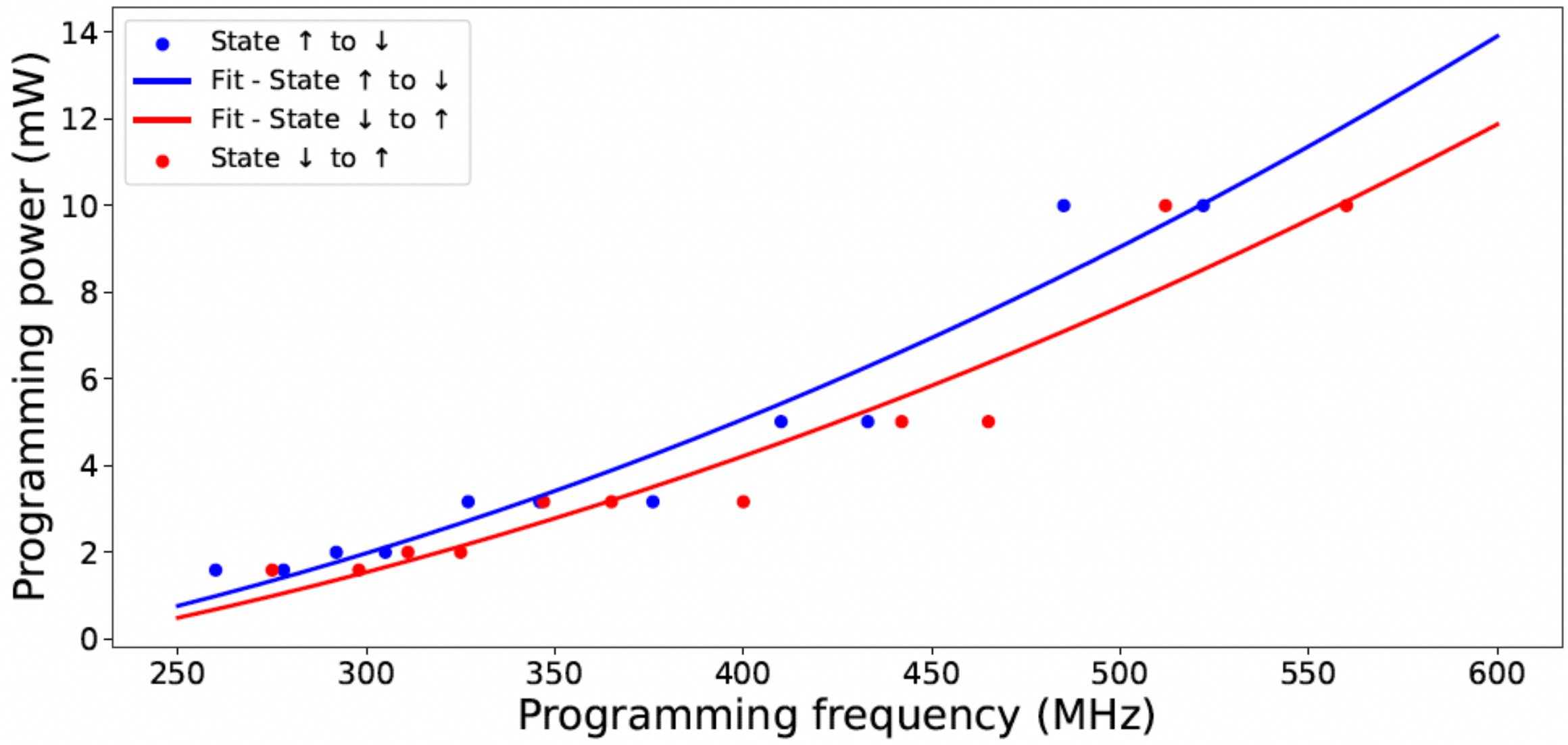

*Figure S 12: Programming power versus programming frequency. Experimental data are represented by red dots (↓ to ↑) and blue dots (↑ to ↓). Quadratic fit associated to data points are plotted in solid lines.*

For resonant pulses, the characteristic duration of switching is set by the build-up and relaxation time of the gyrotropic motion, which corresponds to a few gyration cycles[55] and thus scales as $\tau \propto \frac{1}{f_G}$.

The corresponding pulse energy therefore scales with the frequency:

$$E_{sw} \simeq P_{sw} \times \tau \propto f_G$$

This scaling explains why higher-frequency junctions require larger write power in the present experiments, while lower-frequency junctions are expected to require longer but energetically more favorable resonant pulses[41,53,54,56,57].

## Benchmarking with other in-memory computing devices

Table 3, adapted from Lanza et al.[2], compares our results to other in-memory computing technologies for programming speed and energy. We observe that the vortex-based RF MTJs are competitive with the other technologies.

*Table 3: Programming speed and energy of the main in-memory computing technologies*

| **Material** | **Programming speed** | **Programming energy** |
|---|---|---|
| **Phase Change Memory** | 20~100 ns [58] | ~10 pJ [59] |
| **Resistive RAM** | >20 ns[60] | ~ 1 pJ[60] |
| **Magnetic RAM** | 50 ns[61] | 12 pJ[61] |
| **Vortex RF MTJ (this work)** | 25 ns | 31 pJ |

## Scaling the number of MTJs in the chain

The possibility to selectively program individual MTJs in a chain relies on each MTJ and each polarity have distinct switching frequency windows.

- The size of a switching window for each MTJ in each polarity is $\Delta F \approx \alpha \times f_{res}$. With $\alpha$ the magnetic damping and $f_{res}$ the resonance frequency.

- We take the conservative approach to allow no overlap between the windows of the two polarities of an MJ, and no overlap between the windows of two MTJs.

In consequence, the frequency of the n-th MTJ should be such as:

$$f_n = f_{n-1} + 2 \times \alpha \times f_{n-1}$$

The result is a geometric sequence of expression $f_n = f_0 \times (1 + 2 \times \alpha)^{n-1}$.

Supposing a frequency range from $f_0$ = 50 MHz to $f_n$ = 1 GHz, we find that:

- With a magnetic damping of $\alpha$ = 5 %, as measured in the present work, we can fit 32 synapses.

- With a magnetic damping of α = 2 %, we can fit 77 synapses.

- With a magnetic damping of α = 1 %, we can fit 152 synapses.

Damping values of 1 to 2 % are realistic for MTJs. In consequence, we estimate that we can build chains of at least 50 to 100 MTJs.